\documentclass[article]{IEEEtran}
\usepackage{amsfonts,amsmath,amssymb,epsfig,graphicx,latexsym,multicol}
\usepackage{times}
\usepackage[usenames,dvipsnames]{pstricks}
\usepackage{epsfig}
\usepackage{pst-grad} 
\usepackage{pst-plot} 
\usepackage{epstopdf}
\usepackage{authblk}

\newtheorem{theorem}{Theorem}
\newtheorem{lemma}[theorem]{Lemma}

\newtheorem{remark}{Remark}

\newcommand\qed{\hfill \rule{1.2mm}{2.8mm}}

\newcommand{\bB}{{\bf B}}
\newcommand{\bc}{{\bf c}}

\newcommand{\bg}{{\bf g}}
\newcommand{\bh}{{\bf h}}

\newcommand{\bq}{{\bf q}}
\newcommand{\br}{{\bf r}}
\newcommand{\bs}{{\bf s}}

\newcommand{\bu}{{\bf u}}
\newcommand{\bv}{{\bf v}}
\newcommand{\bw}{{\bf w}}\newcommand{\bW}{{\bf W}}
\newcommand{\x}{{\bf x}} 
\newcommand{\by}{{\bf y}}\newcommand{\bY}{{\bf Y}}
\newcommand{\bz}{{\bf z}}

\newcommand{\kl}{^{[kl]}}\newcommand{\kld}{^{[kl]^\dag}}
\newcommand{\kj}{^{[kj]}}\newcommand{\kjd}{^{[kj]^\dag}}
\newcommand{\ki}{^{[ki]}}\newcommand{\kid}{^{[ki]^\dag}}
\newcommand{\ks}{^{[ks]}}\newcommand{\ksd}{^{[ks]^\dag}}
\newcommand{\kv}{^{[kv]}}\newcommand{\kvd}{^{[kv]^\dag}}
\newcommand{\nj}{^{[nj]}}
\newcommand{\ns}{^{[ns]}}

\newcommand{\nv}{^{[nv]}}
\newcommand{\nl}{^{[nl]}}\newcommand{\nld}{^{[nl]^\dag}}
\newcommand{\klt}{^{[kl]^2}}
\newcommand{\kjt}{^{[kj]^2}}

\newcommand{\kvt}{^{[kv]^2}}
\newcommand{\njt}{^{[nj]^2}}

\newcommand{\sumvL}{\sum_{v=1}^L}
\newcommand{\sumjL}{\sum_{j=1}^L}
\newcommand{\sumiL}{\sum_{i=1}^L}
\newcommand{\sumsL}{\sum_{s=1}^L}

\newcommand{\sumlL}{\sum_{l=1}^L}
\newcommand{\sumnK}{\sum_{n=1}^K}

 \newcommand{\mbf}[1]{\mathbf{#1}}

\begin{document}
%
\title{Interference Reduction in Multi-Cell Massive MIMO Systems I: Large-Scale Fading Precoding and Decoding}


\author[1]{Alexei Ashikhmin}
\author[1]{Thomas L. Marzetta}
\author[2]{Liangbin Li}
\affil[1]{Bell Laboratories Alcatel-Lucent, 600 Mountain Ave, Murray Hill, NJ 07974.}
\affil[2]{University of California, Irvine, CA 92617.}

\maketitle

\begin{abstract}
A wireless massive MIMO system entails a large number (tens or hundreds) of base station antennas
serving a much smaller
number of users, with large gains in spectral-efficiency and
energy-efficiency compared with conventional MIMO technology.
Until recently it was believed that in multi-cellular massive MIMO system, even
in the asymptotic regime, as the number of service antennas tends to infinity, the performance is limited
by directed inter-cellular interference. This interference results from unavoidable re-use of
reverse-link training sequences (pilot contamination) by users in different cells.

We devise a new concept that leads to the effective elimination of inter-cell interference in massive MIMO systems.
This is achieved by outer multi-cellular precoding, which we call  Large-Scale Fading Precoding (LSFP).
The main idea of LSFP is
that each base station linearly combines messages aimed to users from different cells that re-use the same training  sequence.
Crucially, the combining coefficients depend only on the {\em slow-fading
coefficients} between the users and the base stations.
Each base station independently transmits its LSFP-combined symbols using
conventional linear precoding that is based on estimated {\em fast-fading
coefficients}. Further, we derive estimates for downlink and uplink SINRs and
  capacity lower bounds for the case of massive MIMO systems with LSFP and a finite number of base station antennas.

\end{abstract}

\section{Introduction}

Multiple-input multiple-output (MIMO) technology has
been a subject of intensive studies
during the last two decades. This technology
became a part of many wireless standards since it can significantly improve
the efficiency  and reliability of wireless systems. Initially, research in this area
was focused on the point-to-point communication scenario, when two devices are equipped
with multiple antennas communicate to each other. In recent years,
 the focus has shifted to multi-user multiple-input multiple-output (MU-MIMO) systems, in which a base station
 is equipped with multiple antennas and simultaneously serves a multiplicity of autonomous
  users. These users can be cheap single-antenna devices and most of the expensive equipment
 is needed only at base stations.
 Another advantage of MU-MIMO systems is their high multi-user diversity, which allows making
 the system performance more robust to the propagation environment than in the case of point-to-point
 MIMO case.
   As a result, MU-MIMO has
become an integral part of communications standards, such as
802.11 (WiFi), 802.16 (WiMAX), LTE, and is progressively
being deployed throughout the world.

In most modern MU-MIMO systems, base stations have only a few, typically fewer  than 10, antennas, which
results in relatively modest spectral efficiency and precludes a rapid increase
in data rates, as well as higher user density required in the next generation cellular networks. This, along
with the GreenTouch initiative to decrease the power consumption in communication networks, motivated
extensive research on Massive MIMO systems, where each base station is equipped with a significantly larger
 number of antennas, e.g., 64 or more.

 In \cite{Marzetta10} and \cite{Marzetta11}, Marzetta used asymptotic arguments based on
random matrix theory to show that the effects of additive noise
and intra-cell interference, and
the required transmitted energy per bit vanishes as the number
of base station antennas grows to infinity. Furthermore,
simple linear signal processing approaches, such as matched filter
precoding/detection, can be used to achieve  these advantages.

Another important advantage of massive MIMO system is their energy
efficiency. In \cite{Ngo}, it is shown that the transmit
power scales down linearly with the number of
antennas. 
The high energy efficiency of massive MIMO systems is very important
since the large energy consumption can be one of the main technical issues
in future wireless networks \cite{Xu}, \cite{Xion}.  

Because of the above advantages, in recent years, massive MIMO systems attracted
significant attention of the research community. Understanding signal
processing, information theoretic properties, optimization of parameters, and other
aspects of massive
MIMO system become subjects of intensive studies. The articles \cite{Rusek}, \cite{Larsson}, and \cite{Lu}
present a good introduction into this area,
including fundamental
information theoretical limits, antenna and propagation aspects, design of precoder/decoder,
and other technical issues.

In this work, we consider  one of the most important and interesting problems of massive MIMO systems - pilot contamination.
In \cite{Marzetta11} (see also \cite{Fernandec} and\cite{Fernandec1}), Marzetta derived estimates for SINR values in a non-cooperative cellular network in the asymptotic regime when the number of base station antennas tends to infinity. He showed that in this regime not all interference vanishes, and therefore, SINR does not grow indefinitely. The reason for this is that, unless all users in the network use mutually orthogonal training sequences (pilots), the training sequences  transmitted by different users contaminate each other. As a result,
 the estimates of channel state information  made at a base station are biased toward 
 users located in other cells. This effect is called {\em pilot contamination  problem}. The pilot contamination causes 
 the inter-cell interference that is proportional to the number of base station antennas.

 A number of techniques were proposed for mitigation the pilot contamination. 
The numerical results obtained in \cite{Fernandec},\cite{Fernandec1}, and \cite{hoydis} show that these techniques provide
 breakthrough data transmission rates for noncooperative cellular networks. Advanced network MIMO systems that allow some collaboration between base stations were proposed recently in \cite{huh}. Unfortunately, in all these
techniques SINR values remain finite and do not grow indefinitely with the number of base station antennas $M$.

In \cite{patent} and \cite{Ashikhmin2012},  the authors proposed massive MIMO systems with limited collaboration between base stations and an outer multi-cellular precoding. This outer-cellular precoding is based only on the use of large-scale fading
coefficients betweens base stations and users. Since large-scale fading coefficients do not depend on
antenna index and frequency, the number of them is relatively small - only one coefficient for each pair of a base station and an user. In addition, these coefficients change slowly over time. Thus, the proposed outer-cellular precoding does not require extensive
traffic between base stations and/or a network controller.
In the asymptotic
regime, as $M$ tends to infinity, this outer-cellular  precoding allows one to construct interference and noise free multi-cell massive MIMO  systems with frequency reuse one. In this work, we present these results in full details with rigorous proofs and extend them to the real life scenario when the number of base station antennas is finite.

In \cite{patent} and \cite{Ashikhmin2012}, the main goal for designing the outer-cellular precoding was cancelation of  the interference caused by the pilot contamination, and therefore we called it
{\em pilot contamination precoding}. It happened, however, that in the regime of a finite number of antennas, other
sources of interference, caused not by pilot contamination, can not be ignored. The outer-cellular
 precoding allows one to efficiently mitigate these sources of interference. For this reason, we decided that {\em large-scale fading precoding} is a better name for it.

The paper is organized as follows. First, in Section \ref{sec:model}, we describe our system model, network assumptions, and TDD protocol.
Then, in Section \ref{subsec:Pilot Contamination}, we explain the pilot contamination problem and why inter-cell
 interference does not vanish as the number of base station antennas grows. In Sections \ref{subsec:DPCP} and \ref{subsec:UPCP} we propose large-scale fading precoding and decoding protocols and show that in the asymptotic regime,  as the number of base station antennas tends to infinity, these protocols allow construction of  interference and noise free massive MIMO systems. In Section \ref{subsec:Estimation of Large-Scale Coef},
 we briefly outline an approach for estimation of large-scale fading coefficients.
 In Section \ref{sec:finiteM}, we analyze LSFP and LSFD in the regime of a finite number of base station antennas. In particular,
 in Section \ref{subsec:LSFP with fininite M} and \ref{subsec:LSFD with fininte M},
 we derive estimates for downlink and uplink SINRs and the capacity of massive MIMO systems with large-scale precoding and decoding respectively. Further, in \ref{subsec:NoLSFP and ZF-LSFP}, we consider two cases:  when LSFP is not used, and when Zero-Forcing LSFP is used.  In \ref{subsec:optimization problem},  we formulate an optimization problem
 for designing efficient LSFP and in \ref{subsec:first simul results}, we present simulation results for Zero-Forcing
 LSFP for different number of base station antennas.


\section{System Model}\label{sec:model}

We consider a two-dimensional hexagonal cellular network composed of  $L$ hexagonal cells with one base station and $K$ users in each cell.
Each base station is equipped with $M$ omnidirectional antennas and each user has a single omnidirectional antenna.   We further assume that Orthogonal Frequency-Division Multiplexing (OFDM) is used and that in a given
subcarrier we have a flat-fading channel.

   The throughput of 95\% of users can be improved by using massive MIMO system with frequency reuse factor greater than one \cite{Marzetta11}.
   Alternatively one can improve the throughput by using a pilot reuse factor greater than one \cite{Hong2013}. For instance, a pilot reuse factor of seven allows assigning orthogonal pilots to users in two concentric rings of cells, at the expense of a longer training time. Similarly, the results presented in this work can be further improved by applying a frequency reuse factor or  pilot reuse factor different from one. However, to make presentation shorter, we will consider only the case of frequency reuse and pilot reuse one. Thus the entire frequency band is used for downlink and uplink transmissions by all base stations and all users, and the length of the pilot sequences is equal to the number of users in one cell.

\subsection{Channel Model}\label{subsec:channel_model}
\begin{figure}[htb]
\centering
\includegraphics[scale=0.35]{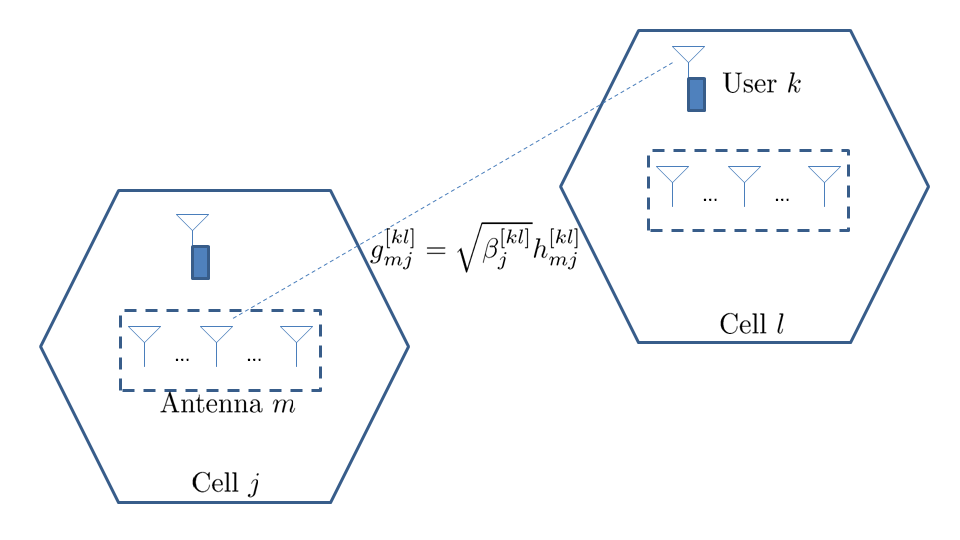} 
\caption{The channel coefficient $g_{mj}^{[kl]}$ between the $m$-th antenna of the $j$-th cell and the $k$-th terminal in the $l$-th cell}
\label{fig:cells}
\end{figure}

For a given subcarrier, we denote by
 \begin{equation}\label{eq:g}
  g_{mj}^{[kl]}= \sqrt{\beta_{j}^{[kl]}}h_{mj}^{[kl]}
 \end{equation}
  the {\em channel (propagation) coefficient}  between the $m$-th antenna of the $j$-th base station and the $k$-th terminal  of the $l$-th cell, Fig.~\ref{fig:cells}.
  The first factor in (\ref{eq:g}) is the large-scale fading coefficient $\beta_{j}^{[kl]}\in \mathbb{R}^+$ and the second factor
  is the small-scale fading coefficient $h_{mj}^{[kl]}\in {\mathbb C}$.

A  large-scale fading coefficient depends on the shadowing and distance between the corresponding user and base station. Typically, the distance between a user and base station is significantly larger than the distance between
  base station antennas. For this reason, the standard assumption is  that the large-scale fading coefficients do not depend on the antenna index $m$ of a given base station.
  We also assume that these coefficients are constant across the used frequency band, i.e., that they do not depend on
  OFDM subcarrier index.
  Thus there is only one large-scale fading coefficients for each pair of a user and a base station.     A detailed model for
 the large-scale fading coefficients will be given in Part II of the paper \cite{Part II}.

  In contrast, the small-scale fading coefficients depend on both the antenna and subcarrier indices. Hence, if $N$ is the number of OFDM subcarriers, then for each pair of a user and base station there are about $NM$ small-scale fading coefficients.
  In what follows, we consider only one OFDM subcarrier and so we do not write the subcarrier index for small-scale fading coefficients. For small-scale fading coefficients, we assume Rayleigh fading model, i.e., $h_{mj}^{[kl]}\sim {\cal CN}(0,1)$ and for any $(m,j,k,l)\not =(u,i,n,v)$  the coefficients $h_{mj}^{[kl]}$ and $h_{ui}^{[nv]}$ are independent. From the above assumptions it follows that for any $(j,k,l)\not (i,n,v)$ the vectors $\bh_j\kl$ and
  $\bh_i^{[nv]}$ (similarly $\bg_j\kl$ and $\bg_i^{[nv]}$) are independent.

The small-scale fading coefficients between the $j$-th base station  and the $k$-th user in the $l$-th cell form {\em small-scale fading vector}
  $$
  \mbf h_{j}^{[kl]}=\left(h_{1j}^{[kl]},h_{2j}^{[kl]},\ldots,h_{Mj}^{[kl]}\right)^T\in \mathbb C^{M\times 1}.
  $$
The channel coefficients  between the $j$-th base station and the $k$-th user in the $l$-th cell form {\em channel vector}
  $$
  \mbf g_{j}^{[kl]}=\left(g_{1j}^{[kl]},g_{2j}^{[kl]},\ldots,g_{Mj}^{[kl]}\right)^T=
  \sqrt{\beta_j^{[kl]}} \mbf h_j^{[kl]}\in \mathbb C^{M\times 1}.
  $$

Since small-scale fading coefficients are i.~i.~d., we have $\mbf h_{j}^{[kl]}\sim {\cal CN}(0,\mbf I_M)$
and $\mbf g_{j}^{[kl]}\sim {\cal CN}(0,\beta_j\kl \mbf I_M)$.

We further assume a block fading model, that is,  that small-scale coefficients $h_{j}^{[kl]}$ stay constant during coherence blocks of $T$ OFDM symbols.  The
small-scale fading coefficients in different coherence blocks are assumed to be independent.
Similarly, we assume that large-scale fading coefficients $\beta_j\kl$ stay constant during large-scale coherence blocks of $T_\beta$ OFDM symbols. Typically $T_\beta$ is significantly larger than $T$
 (at the end of Section  \ref{sec:PCP}  we discuss this in more details).
 For different large-scale coherence blocks coefficients $\beta_j\kl$ are assumed being independent.

Since large-scale fading coefficients stay constant for long coherence blocks and the number of these coefficients is relatively small (for each pair of base station and mobile there is only one large-scale fading coefficient) throughout the paper we use the following

\centerline{\em Network Assumptions I:}

\begin{enumerate}
\item
 We assume that the $j$-th base station can accurately estimate and track
 large-scale fading coefficients $\beta_{j}^{[kl]}$ with $k=1,K$ and $l=1,L$.
\item If $\epsilon_j\kj$ is a quantity that depends only on large-scale fading coefficients we assume that the $j$-th base station has means to forward it to the $k$-th user in the $j$-th cell.
\end{enumerate}
Throughout the paper,
in our analysis of communication protocols we will not take into account the resources needed
for implementing the above assumptions.

Finally, we assume reciprocity between uplink and downlink channels, i.e., $\beta_{j}^{[kl]}$ and $\mbf h_{mj}^{[kl]}$ are the same for these channels. The reciprocity, up to high accuracy, can be achieved with proper calibration of
hardware components, e.g.,  the transmitted power amplifier and the received low-noise amplifier.

\subsection{Time-Division Duplexing Protocol}\label{subsec:TDD Protocol}

We consider a wireless network, where each cell has $K$ users enumerated by index $k$ from $1$ to $K$.
In each cell, the same set of $K$ orthonormal training sequences ${\br}^{[k]}\in \mathbb{C}^{\tau\times 1}$ ( $\br^{[k]^\dag}\br^{[i]}=\delta_{ki}$) are assigned to the users. The sequence $\br^{[k]}$ is assigned to the $k$-th user. Since small-scale fading vectors are mutually independent for different coherence blocks, we have to assume that $\tau <T$.
Since the number of orthogonal $\tau$-tuples can not exceed $\tau$, we also have $K\le \tau$.

The assumption that the same set of training sequences is used in all cells is justified in the following.
If users move fast, the coherence block is short, that is,  $T$ is small. Hence $\tau$, the training time, should be also small.
Therefore, it is reasonable to assume that we can assign orthogonal training sequences to users within one cell,
but there are not enough orthogonal training sequences for users from different cells. Thus, we
have to reuse the same training sequences in all cells.

There is an alternative scheme in which different sets
of training sequences are used in different cells. Thus, in the $l$-th cell the orthonormal sequences $\br_{l}^{[k]}\in {\mathbb C}^{\tau\times 1},k=1, K$, are used. In this case, however, the training sequences from different cells are still nonorthogonal, that is, for generic $\br_{l}^{[k]}$ and
$\br_{n}^{[i]}$, we have $\left|\br_{l}^{[k]^\dag}\br_{n}^{[i]}\right|>0$. Our estimates show that such a scheme would achieve performance similar
to the scheme with the same set of training sequences in all cells. At the same time, the analysis becomes more complex. For this reason, in this paper, we do not consider the alternative scheme, which, however, could be done in future works.

\begin{figure}[htb]
\centering
\includegraphics[scale=0.3]{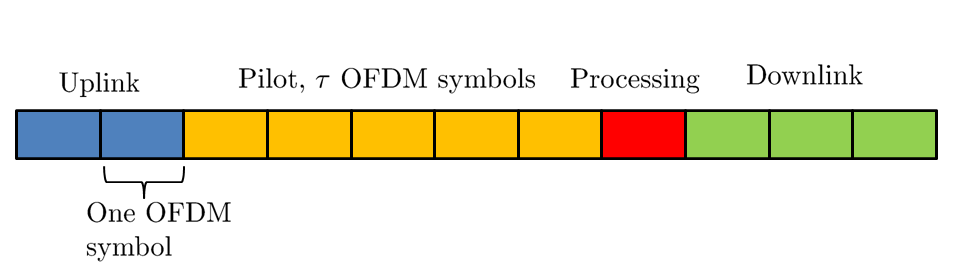} 
\caption{Coherence block of $T=11$ OFDM symbols}
\label{fig:TDDprotocol}
\end{figure}

The Time-Division Duplexing (TDD) protocol consists of six steps. The last four steps, that are conducted during
 each coherence block, are shown in Fig.\ref{fig:TDDprotocol}.

\noindent{\bf Time-Division Duplexing Protocol}
\begin{enumerate}
\item In the beginning of each large-scale coherence block (of duration $T_\beta$ OFDM symbols)
the $j$-th base station estimates the large-scale fading coefficients
$\beta_j^{[kl]},k=1,K,\;l=1,L$.
\item Next, the $j$-th base station transmits to $K$ mobiles located in the $j$-th cell
the quantities
$$
\epsilon_j^{[kj]}=={\sqrt{M \rho_f \rho_r\tau}\beta\kj_j\over (1+\sumsL \rho_r\tau \beta\ks_j)^{1/2}},~k=1,K.
$$
\item All users synchronously transmit their uplink signals $x\kj,\;k=1,K,\;j=1,L$.
 \item All users synchronously transmit training sequences $\br^{[k]}$.
 \item Base stations process received uplink signals and training sequences.
  In particular, each base station estimates the channel vector between itself and the users located within the same cell, and further performs decoding and precoding of uplink and downlink signals, respectively.
 \item All base stations synchronously transmit their downlink signals $\x_j,j=1,L$.
 \end{enumerate}

\noindent{\bf The End}

In what follows, we consider these steps in details.
It is convenient to start with Step 3.
{\bf During Step 3} of the TDD protocol, the $j$-th base station receives uplink data signal of the form
\begin{equation}\label{eq:yj}
\by_j=\sqrt{\rho_r} \sum_{k=1}^K \sum_{l=1}^L \bg_j^{[kl]}x^{[kl]}+\bw_j\in \mathbb{C}^{M\times 1},
\end{equation}
where $x^{[kl]}$ is the uplink signal of the $k$-th user located in the $l$-th cell, $\rho_r$ is the reverse link transmit power, and
$\bw_j\sim {\cal CN}(0,\mbf I_M)$ is the additive noise.  We assume that all users have the same transmit power.

  {\bf In Step 4,} the $j$-th base station receives training signals, which can be written into a matrix $\bY_j\in \mathbb{C}^{M\times \tau}$ of the form
$$
\bY_j=\sqrt{\rho_r\tau } \sum_{k=1}^K\sum_{l=1}^L \bg_j^{[kl]} \br^{[k]^\dag} + \bW_j,
$$
where $\bW_j\in \mathbb{C}^{M\times \tau}$ is the additive white Gaussian noise matrix with i.i.d. ${\cal CN}(0,1)$ entries.

{\bf In Step 5,} the $j$-th base station uses the fact that the training sequences are orthogonal to obtain the MMSE estimate
of the channel vectors $\bg_{j}^{[kl]}$ as (see for example \cite[Chapter 12]{Kay})
\begin{equation}\label{eq:hat g}
\hat{\bg}_j^{[kj]}=\bY_j\left(\theta_j^{[kj]}\br^{[k]}\right)=\theta_{j}^{[kj]}
\sum_{s=1}^L \sqrt{\rho_r\tau} \bg_j^{[ks]}+\hat{\bw}\kj_j.
\end{equation}
where
$$
\theta_j^{[kj]}={\sqrt{\rho_r\tau}\beta_j^{[kj]}\over 1+\sum_{s=1}^L \rho_r\tau \beta_j^{[ks]}},\mbox{ and } \hat{\bw}_j\sim {\cal CN}(0,\theta_j\kjt \mbf I_M).
$$
According to our Network Assumptions, base station have access to $\beta_j^{[ks]}$ and therefore
are capable of finding $\hat{\bg}_j^{[kj]}$.

Note that the vector $\hat{\bg}_j^{[kj]}$ has the distribution
\begin{equation}\label{eq:dist of g_j^kj}
\hat{\bg}_j^{[kj]} \sim {\cal CN}\left(0,{\rho_r\tau\beta_j^{[kj]^2}\over 1+\sumsL \rho_r\tau\beta_j^{[ks]}}I_M\right).
\end{equation}

Further, the $j$-th base station uses the estimates $\hat{\bg}_j^{[kj]}$ for decoding the transmitted uplink signals $x^{[kj]},k=1,K$,
and forming precoding beamforming vectors for downlink transmission. The decoding and precoding can be implemented in several
possible ways, e.g., zero-forcing, MMSE, or matched filtering. In this work we always assume that the base station uses
matched filtering for decoding and conjugate precoding for downlink transmission. Thus, it gets an estimate of the uplink signal $x^{[kj]}$ as
\begin{align}\label{eq:MF_uplink}
\hat{x}^{[kj]}&=\hat{\bg}_j^{[kj]^\dag}\by_j \nonumber \\
&=\sqrt{\rho_r}
\hat{\bg}_j^{[kj]^\dag}
 \bg\kj_j x\kj + \sum_{n=1 \atop n\not =k}^K \sqrt{\rho_r}\hat{\bg}_j^{[kj]}\bg_j^{[nj]}x^{[nj]}
 \nonumber\\
 &+\sum_{{l=1 \atop l\not =j}}^L \sum_{n=1}^K \sqrt{\rho_r}  \hat{{\bg}}_j^{[kj]^\dag} \bg_j^{[nl]}x^{[nl]} +\hat{{\bg}}_j^{[kj]^\dag}w^{[kj]}.
\end{align}
 For the downlink transmission, the $j$-th base station forms conjugate precoding beamforming vectors as
\begin{equation} \label{eq:conjugat beamforming}
\bu_j^{[k]}={\hat{\bg}_j^{[kj]^\dagger}\over
\lambda_j^{[kj]}
},~k=1, K,
\end{equation}
where  $\lambda_j^{[kj]^2}=\mathbb{E}[||\bg\kj_j||^2]$ is the normalization factor, which according to
\eqref{eq:dist of g_j^kj}, is equal to
\begin{equation}\label{eq:lambda_j^kj}
\lambda_j^{[kj]^2}=M\cdot {\rho_r\tau\beta_j^{[kj]^2}\over 1+\sumsL \rho_r\tau\beta_j^{[ks]}}.
\end{equation}
The $j$-th base station next forms the $M$-dimensional vector
$$
\x_j=\sqrt{\rho_f}\sum_{k=1}^K \bu_j^{[k]}s^{[kj]},
$$
where $s^{[kj]}$ is the downlink data signal intended for the $k$-th user located in the $j$-th cell, and $\rho_f$ denotes forward link transmit power used by the $j$-th base station. We assume that all base stations use the same transmit power.

{\bf Finally, in Step 6,}  the $j$-th base station transmits
and from its $M$ antennas the vector $\x_j$.

The $k$-th mobile in the $j$-th cell receives the signal
\begin{align}
\label{eq:y^kj}
y^{[kj]}&=\sum_{l=1}^L \sqrt{\rho_f}\x_l \bg_l^{[kj]} +w^{[kj]}\nonumber\\
&= \sqrt{\rho_f}{\hat{\bg}_j^{[kj]^\dag}\over \lambda_j^{[kj]}}
 \bg\kj_j s\kj + \sum_{n=1 \atop n\not =k}^K \sqrt{\rho_f}{\hat{\bg}_j^{[nj]^\dag}\over \lambda_j^{[nj]}} s^{[nj]}\bg_j^{[kj]}
 \nonumber\\
 &+\sum_{{l=1 \atop l\not =j}}^L \sum_{n=1}^K \sqrt{\rho_f}  {\hat{\bg}_l^{[nl]^\dag}\over \lambda_l^{[nl]}} s^{[nl]}\bg_l^{[kj]} +w^{[kj]}.
\end{align}

Note that the {\em effective channel} ${\hat{\bg}_j^{[kj]^\dag}\over \lambda_j^{[kj]}}
 \bg\kj_j$, in average, has zero phase.
The quantity
$$
\epsilon_j^{[kj]}=\mathbb{E}[{\hat{\bg}\kj_j\over \lambda\kj_j} \bg\kj_j]={\sqrt{M \rho_f \rho_r\tau}\beta\kj_j\over (1+\sumsL \rho_r\tau \beta\ks_j)^{1/2}},~k=1,K,
$$
which the $j$-th base station transmits to $k$-th user
tells the user the expected value of the power of the effective channel. So the user can use the following simple detector
$$
\hat{s}\kj={y\kj\over \sqrt{\rho_f}\epsilon_j\kj}
$$
for estimating the signal $s\kj$. Alternatively,
instead of transmitting $\epsilon_j^{[kj]}$ at step 4 of the TDD protocol,
base stations can send downlink training sequences that would allow users to estimate
their effective channels. We do not elaborate on these possible details of the TDD protocol.

Note also, Steps 1 and 2 should be conducted only one time during each large-scale coherence block. As we noted in Section \ref{subsec:channel_model}, we do not take into account the resources needed for this.

\section{Large-Scale Fading Precoding and Interference Free LSAS}\label{sec:PCP}

\subsection{Pilot Contamination}\label{subsec:Pilot Contamination}

From \eqref{eq:MF_uplink} and \eqref{eq:y^kj} it follows that the uplink and downlink SINRs
can be written in the form presented at the top of the next page
 \begin{figure*}[htb]
\normalsize
\small{
$$
\mbox{SINR}_U\kj
={\rho_r |\hat{\bg}_j^{[kj]^\dag}\bg\kj_j|^2 \over
\sum_{n=1 \atop n\not =k}^K \rho_r|\hat{\bg}_j^{[kj]}\bg_j^{[nj]}|^2
 +\sum_{{l=1 \atop l\not =j}}^L \sum_{n=1}^K \rho_r  |\hat{{\bg}}_j^{[kj]^\dag} \bg_j^{[nl]}|^2 +\mbox{Var}[\hat{{\bg}}_j^{[kj]^\dag}w^{[kj]}]}
$$}
\hrulefill
\vspace*{2pt}
\end{figure*}

\begin{figure*}[htb]
\normalsize
\small{
$$
\mbox{SINR}_D\kj
=
{{\rho_f\over \lambda_j^{[kj]^2}}
|\hat{\bg}_j^{[kj]^\dag}
 \bg\kj_j |^2\over
 \sum_{n=1 \atop n\not =k}^K {\rho_f  \over \lambda_j^{[nj]^2}}
 |\hat{\bg}_j^{[nj]^\dag} \bg_j^{[kj]}|^2
 +\sum_{{l=1 \atop l\not =j}}^L \sum_{n=1}^K {\rho_f \over \lambda_l^{[nl]^2}}
    |\hat{\bg}_l^{[nl]^\dag}
 \bg_l^{[kj]}|^2 +\mbox{Var}[w^{[kj]}]
 }
$$}
\hrulefill
\vspace*{2pt}
\end{figure*}
In \cite{Marzetta11} and \cite{Fernandec},\cite{Fernandec1} the authors considered LSASs in the asymptotic regime when the number of base station antennas $M$ tends to infinity. The following results were obtain.
 \begin{theorem} \label{thm:SINR}
The downlink SINR of the $k$-th terminal in the $j$-th cell for precoding (\ref{eq:conjugat beamforming}) converges to the following limit:
\begin{equation} \label{eq:SINR1}
 \lim_{M\to\infty}  \mbox{SINR}_D^{[kj]} \stackrel{\textrm{a.s.}}{=} {\beta_{j}^{[kj]^2}/\eta_{j}^{[k]^2} \over
  \sum_{l=1,\atop l\not =j}^L \beta_{l}^{[kj]^2}/ \eta_l^{[k]^2}},
\end{equation}
where
 $$
\eta_l^{[k]}=\left(1+\sum_{s=1}^L \rho_r\tau\beta_l^{[ks]}\right)^{1/2}.
$$
\end{theorem}
To give an intuitive explanation of this result we remind that according to the strong law of
large numbers we have the following lemma.
\begin{lemma}\label{lem:LargeNumbers}
Let $\mbf x,\mbf y \in \mathbb C^{M \times 1}$ be two independent  vectors with distribution $\mathcal{CN}\left(\mbf 0,\nu\,\mbf I_M\right)$. Then
\begin{equation}\label{indlim}
\lim_{M\to\infty} \frac{\mbf x^\dagger\,\mbf x}{M}\stackrel{\textrm{a.s.}}{=}\nu \mbox{ and }
\lim_{M\to\infty} \frac{\mbf x^\dagger\,\mbf y}{M} \stackrel{\textrm{a.s.}}{=} 0.
\end{equation}
\end{lemma}
The signal $\by\kj$ received by the $k$-th user in the $j$-th cell is defined in \eqref{eq:y^kj}
and estimates $\hat{\bg}_l\nl$ are defined in \eqref{eq:hat g}.
It is easy to see that all terms of $\hat{\bg}_l\nl,n\not =k$, are independent from $\bg_l\kj$ and therefore, according to Lemma \ref{lem:LargeNumbers} we have
$$
\lim_{M\rightarrow \infty} {1\over M}\hat{\bg}_l\nld\bg_l\kj \stackrel{\textrm{a.s.}}{=}0.
$$
Hence the contribution into interference of the product $\hat{\bg}_l\nld\bg_l\kj$ vanishes as $M$ grows. At the same time
the estimate $\hat{\bg}_l\kld$ contains $\bg_l\kj$ as a term. The reason for this is that
the $k$-th users in cells $j$ and $l$ are using the same training sequence $\br^{[k]}$. Thus
 the product
$$
{1\over M}\hat{\bg}_l\kld\bg_l\kj
$$
almost surely converges to some finite value and therefore it makes  revanishing contribution  to the interference. This effect is called {\em pilot contamination}. It is shown in \cite{Marzetta11} and \cite{Fernandec},\cite{Fernandec1} that pilot contamination is the only source of interference in the regime of infinitely large $M$.

Similar analysis of the uplink transmission leads to the following result.
\begin{theorem} \label{thm:U_SINR}
The uplink SINR of the $k$-th terminal in the $j$-th cell for decoding (\ref{eq:MF_uplink}) converges to the following limit:
\begin{equation}\label{eq:uplinkSINR}
 \lim_{M\to\infty}  \mbox{SINR}_U^{[kj]} \stackrel{\textrm{a.s.}}{=}
 {\beta_{j}^{[kj]^2}
 \over
 \sum_{l=1,l\not =j}^L
 \beta_{j}^{[kl]^2}}.
\end{equation}
\end{theorem}

 The natural question is whether
one can mitigate the pilot contamination and further increase SINRs beyond results obtained in Theorems
\ref{thm:SINR} and \ref{thm:U_SINR}. Below we briefly discuss some approaches for this.

 One possibility is to use frequency reuse for avoiding interference between adjacent cells \cite{Marzetta11}.
 Though
 the frequency  reuse  reduces  the
bandwidth, overall it allows increasing SINRs for most of the users.

Another possibility  is to
optimize transmit powers. In this case, the $j$-th base station uses the power $\rho_{f}^{[kj]}$ for transmitting to the $k$-th user
in the $j$-th cell and transmits the vector
$$
\x_j=\sum_{k=1}^K \sqrt{\rho_{f}^{[kj]}}\bu_j^{[k]} s^{[kj]}.
$$
Similarly, in the uplink transmission, the $k$-th user in the $j$-th cell will use the transmit power $\rho_{r}^{[kj]}$. Then, the SINR expressions
(\ref{eq:SINR1}) and (\ref{eq:uplinkSINR}) will be respectively
\begin{align*}
\lim_{M\to\infty}   \mbox{SINR}_D^{[kj]} \stackrel{\textrm{a.s.}}{=} & {\rho_{f}^{[kj]}\beta_{j}^{[kj]^2}/\eta_{j}^{[k]^2} \over
  \sum_{l=1,\atop l\not =j}^L \rho_{f}^{[kl]}\beta_{l}^{[kj]^2}/ \eta_l^{[k]^2}}, \mbox{ and }\\
\lim_{M\to\infty}    \mbox{SINR}_U^{[kj]} \stackrel{\textrm{a.s.}}{=}  &
 {\rho_{r}^{[kj]}\beta_{j}^{[kj]^2}
 \over
 \sum_{l=1 \atop l\not =i}^L \rho_{r}^{[kl]}\beta_{j}^{[kl]^2}}.
\end{align*}

It is also possible to use modified TDD protocol in which users from different cells
transmit pilot sequences asynchronously according to the time-shifted protocol proposed in \cite{Appaih},\cite{Fernandec},\cite{Fernandec1}.   For sufficiently large $M$
the time-shifted protocol gives significant increase in downlink and uplink SINRs.

One more possibility is to replace conjugate precoding \eqref{eq:conjugat beamforming} with
another linear precoding that mitigates the pilot contamination effect \cite{Jose}.

In all the above techniques, however, downlink and uplink SINRs approach some finite limits as $M$ tends to infinity.
In other words, SINRs do not grow with $M$.

To obtain SINRs that grow along with $M$, one may try to use the network MIMO approach (see for example \cite{NetworkMIMO1},\cite{NetworkMIMO2},\cite{NetworkMIMO3} and references within). The network MIMO assumes that
 the $j$-th base station estimates the coefficients
$\beta_j\kl$ and $h_{mj}^{[kl]}$ for all $k=1, K$ and $l=1, L$, and sends them
to other base stations. After that, all base station start to behave as one super large antenna array.
This approach, however, seems to be infeasible for the following reasons.

First, the number of small-scale fading coefficients $h_{mj}^{[kl]}$
 is proportional to $M$. Thus, in the asymptotic regime, as $M$ tends to infinity, the needed traffic between base stations also infinitely grows and the network MIMO becomes infeasible.

 Even in the case of a finite $M$ the needed traffic is tremendously large.
Indeed, the small-scale fading coefficients depend on frequency.
 The typical assumption is that
 small-scale fading coefficients $h_{mj}^{[kl]}$ of OFDM
tones $i$ and $i+\Delta$ are considered being independent random variables when $\Delta$ is the coherent bandwidth, which typically is in between $10$ and $20$ subcarriers..
   Thus, if $M=100$ and the total number of OFDM subcarriers is say
$N=1400$, and $\Delta=14$, then the $j$-th base station needs to transmit to other base stations
$NM/\Delta\cdot K(L-1)=10000K(L-1)$ small-scale fading coefficients for given $k$ and $l$. Note that the small-scale fading coefficients substantially change as soon as a mobile moves a quarter of the wavelength. Taking all this into account we conclude
that the needed traffic between base stations is hardly feasible.

The second reason is even more fundamental. Since users in different cells reuse the same pilot sequences, the $j$-th base station is
not capable of obtaining independent estimates for the coefficients $h_{mj}^{[kj]}$ and $h_{mj}^{[kl]}$, since the $k$-th users in cells $j$ and $l$ use the same pilot sequence $\br^{[k]}$. Thus, the standard network MIMO approach is not applicable even if we
ignore the traffic problem.

One possible conclusion of the above arguments can be that in both noncooperative LSASs and LSASs with cooperation (like network MIMO), SINRs do not grow
with $M$ beyond certain limits. In this paper, we disprove the above statement. We demonstrate that limited cooperation between base stations allows us to completely resolve
the pilot contamination problem and to construct interference and noise free LSASs with infinite downlink and uplink SINRs.

\subsection{Large-Scale Fading Precoding}\label{subsec:DPCP}
We start with changing the Network Assumption defined in Section \ref{sec:model}.

\centerline{\em Network Assumptions II}

\begin{figure}[htb]
\centering
\includegraphics[scale=0.6]{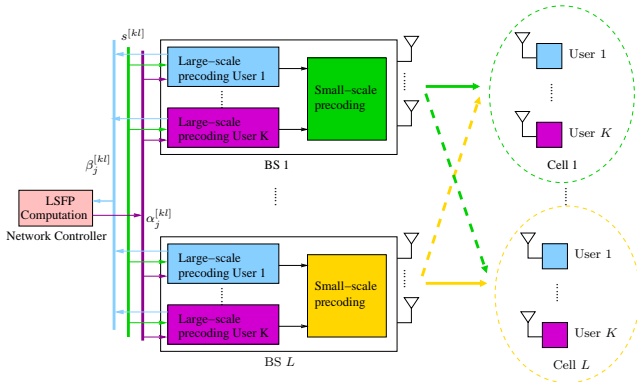}
\caption{System diagram for the LSFP. Each BS performs two levels of precoding. Multi-cell cooperation is based on the large-scale fading coefficients only. Each BS also performs local precoding using estimates of $M$-dimensional fast fading vectors.}
\label{fig:BlockDiag}
\end{figure}

\begin{enumerate}
\item
 We assume that the $j$-th base station can accurately estimate and track
 large-scale fading coefficients $\beta_{j}^{[kl]},~k=1,K$ and $l=1,L$.
 \item If $\epsilon_j\kj$ is a quantity that depends only on large-scale fading coefficients
we assume that the $j$-th base station can forward it to the $k$-th user in the $j$-th cell.
 \item We assume that all base stations are connected to a network controller (as it is shown in
 Fig. \ref{fig:BlockDiag})
 and that the large-scale fading coefficients  $\beta_{j}^{[kl]}$ are accessible to the network controller.
 \item We assume that all downlink signals $s^{[kj]}$ with $j=1,L,\;k=1,K$, are accessible to all base stations.
\end{enumerate}

\begin{remark}
 We assume that the network controller and base stations have access to all, across the entire network,  $\beta_{j}^{[kl]}$ and $s^{[kj]}$ respectively,
only in order to obtain  a simple theoretical model. In Part II of the paper \cite{Part II}, we will replace these assumptions with more realistic ones.
\end{remark}

\begin{remark}
 We would like to point out that the large-scale  fading coefficients $\beta_{j}^{[kl]}$ are relatively easy to estimate and track. One possible approach for this is outlined at the end of this section.
 \end{remark}

Below we describe the {\em Large-Scale Fading Precoding} protocol
for interference mitigation in LSASs.
Originally we designed this protocol in cite \cite{patent} and \cite{Ashikhmin2012} for canceling the directed interference caused by pilot contamination and called it {\em Pilot Contamination Precoding} (PCP). Recently, however,
 we came to the conclusion that
 that the name Large-Scale Fading Precoding better reflects the idea of this protocol.

 \noindent{\bf Large-Scale Fading Precoding (LSFP)}
\begin{enumerate}
\item  In the beginning of each large-scale coherence block (of duration $T_\beta$ OFDM symbols)
the $j$-th base station estimates the large-scale fading coefficients
$\beta_j^{[kl]},k=1,K,\;l=1,L$, and sends them to the network controller.

\item The network controller computes the $L\times L$ LSFP precoding matrices
$$\Phi^{[k]}=\left(\begin{array}{c}
\underline{\phi}_1^{[k]}\\
\underline{\phi}_2^{[k]}\\
\vdots\\
\underline{\phi}_L^{[k]}
\end{array}\right),\; k=1,K,
$$
as functions of $\beta_j^{[kl]},\;j,l=1,\ldots,L$, so that
\begin{equation}\label{eq:||alpha_j||<1}
||\underline{\phi}_l^{[k]}||^2\le 1,
\end{equation}
and sends the rows $\underline{\phi}_j^{[k]},\;k=1,K,$
to the $j$-th base station. \\

\item \label{step:DL PCP epsilon} The network controller computes quantities
$$
\epsilon\kj_j=\sqrt{\rho_f} M \rho_r \tau \sum_{l=1}^L { \beta_l\kj\beta_l\kl\over
1+\sum_{s=1}^L \rho_r \tau\beta_l\ks} {\phi_l\kj\over \lambda_l\kl},\;k=1,K,
$$
(where $\phi\kl_j$ is the $(j,l)$ entry of $\Phi^{[k]}$)
and sends them to the $j$-th base station, which sends them further to the corresponding users located in the $j$-th cell.

\item The $j$-th base station conducts {\bf large-scale fading precoding}. Namely, it computes signals
\begin{equation}\label{eq:tilde s}
{c}_j^{[k]}= \underline{\phi}_j^{[k]} \left(\begin{array}{c}
s^{[k1]}\\
s^{[k2]}\\
\vdots\\
s^{[kL]}
\end{array}\right),\;k=1,K.
\end{equation}
(Since $\mbox{Var}[s\kj]=1$, the constraint \eqref{eq:||alpha_j||<1} implies that $\mbox{Var}[c_j^{[k]}]=1$.)
\item The $j$-th base station obtains the MMSE estimates $\hat{\bg}_j^{[kj]},k=1,K$,
according to (\ref{eq:hat g}).

\item \label{step:DL PCP small-scale} The $j$-th base station performs {\bf small-scale fading precoding}, namely it forms conjugate precoding beamforming vectors
$$
\bu_j^{[k]}={\hat{\bg}_j^{[kj]^\dag}\over \lambda_j\kj}, k=1,K,
$$
 and transmits from its $M$ antennas the vector
$$
\x_j=\sqrt{\rho_f} \sum_{k=1}^K  \bu_j^{[k]} {c}_j^{[k]}.
$$
(Note that other types of small-scale precodings can be used at this step. For instance, vectors
$\bu_j^{[k]}$ can be formed with the help of $M$-dimensional zero-forcing precoding.)
\end{enumerate}
\noindent{\bf The End}
\vspace{0.1cm}

The block diagram of this protocol is shown in Fig.\ref{fig:BlockDiag}.

We would like to point out the following things about this algorithm.
\begin{itemize}
\item No exchange of small-scale fading coefficients $h_{mj}\kl$ between base stations and the network controller is required (opposite to the network MIMO).
\item Steps 1 -- 3 are conducted once every large-scale coherence block,  which is typically about $40$ times longer than the coherence blocks of small-scale coefficients.
\item The purpose of quantities $\epsilon\kj_j$ is to send to the corresponding users the expected powers of their  effective channels (see \eqref{eq:ykl1} and \eqref{eq:numerator} in Section \ref{sec:finiteM} for details). If the $k$-th user in the $j$-th cell receives the signal $y\kj$ it can estimate the signal $s\kj$ as
    $$
    \hat{s}\kj={y\kj\over \epsilon_j\kj}.
    $$
    Alternatively, instead of sending $\epsilon\kj_j$ base stations can send downlink training sequences.
\item The estimates $\hat{\bg}_j^{[kl]}$ are computed one time  for each coherence block of small-scale coefficients, that is once every $T$ OFDM symbols.
\item The quantity $c_j^{[k]}$ and vector $\x_j$
are computed for each channel use (each downlink OFDM symbol).
\end{itemize}

Now we show that an appropriate choice of LSFP precoding matrices $\Phi^{[k]}$ allows one to completely cancel the
interference and noise as $M$ tends to infinity.
Let us define $L\times L$ matrices composed of large scale fading coefficients
\begin{equation}\label{eq:BDk}
\mbf B^{[k]}_\mathrm{D}=\left(\begin{array}{lll}
\beta_{1}^{[k1]}/\eta_1^{[k]}& \ldots & \beta_{L}^{[k1]}/\eta_L^{[k]}\\
\vdots & & \vdots \\
\beta_{1}^{[kL]}/\eta_1^{[k]} & \ldots & \beta_{L}^{[kL]}/\eta_L^{[k]}\end{array}
\right).
\end{equation}
For $B_D^{[k]}$ of full rank we define {\em Zero-Forcing LSFP} (ZF-LSFP) as the above LSFP with
\begin{equation}\label{eq:A}
 \Phi^{[k]}=\sqrt{\rho_A}\bB^{[k]^{-1}}_\mathrm{D},\;k=1, K,
\end{equation}
where $\rho_A$ is a normalization factor to insure the constraint \eqref{eq:||alpha_j||<1}.

For analysis of ZF-LSFP it is convenient to
make a small modification of the LSFP assuming that at Step 3  $\epsilon\kj_j=\sqrt{M\rho_f\rho_r\rho_A\tau}$.

Let us assume now that signals $s\kj$ are taken from a signal constellation ${\cal R}=\{r_1, \ldots, r_N\}$ according to some probability mass function (PMF) $P_S(\cdot)$. Define further the entropy
$$
H(s\kj)=-\sum_{s\in{\cal R}} P_S(s)\log P_S(s).
$$
Assume also that all $|r_j|,r_j\in {\cal R}$, are finite.

The $k$-th user in the $j$-th cell receives the signal
  \begin{equation} \label{eq:y^kj PCP0}
y^{[kj]}=\sumlL \sum_{n=1}^K{\sqrt{\rho_f}\over \lambda\nl_l}
\hat{\bg}_l^{[nl]^\dag}
 \bg_l\nj c_l^{[n]}+w^{[kj]}.
\end{equation}

Let the user uses the following simple detection method of the transmitted signal $s\kj$:
\begin{equation}\label{eq:s\kj_M}
\hat{s}_M\kj={y\kj \over \sqrt{M\rho_f\rho_r\rho_A\tau}}.
\end{equation}

The following theorem shows that in the asymptotic regime, as $M\rightarrow \infty$, the ZF-LSFP allows one to reliably transmit signals from an arbitrary large ${\cal R}$ and therefore it provides infinite capacity.
\begin{theorem}\label{thm:PCP_D_ZF} For ZF-LSFP we have
\begin{equation} \label{SINR_ZF_PCP_downlink}
  \lim_{M\rightarrow \infty} I(\hat{s}_M\kj; s\kj)=H(s\kj),\;k=1,K,\;j=1,L.
  \end{equation}
  \end{theorem}

To prove this theorem we need the following lemma.
 Let
$$
s_M=a_M\cdot s+w_M,
$$
where $s$ is a random signal from ${\cal R}$
and $a_M$ and $w_M$ are random variables (not necessarily  independent) such that
\begin{equation}\label{eq:aM a.s. 1}
\lim_{M\rightarrow\infty} a_M \stackrel{\textrm{a.s.}}{=}1 \mbox{ and } \lim_{M\rightarrow\infty} w_M \stackrel{\textrm{a.s.}}{=}0.
\end{equation}
\begin{lemma}\label{lem:inf_capacity}
$$
\lim_{M\rightarrow\infty} I(s_M; s)= H(s).
$$
\end{lemma}
A proof of this lemma is in Appendix A.

We will also need the following well known fact. Let $\phi$ and $\beta$ be some constants and $a_M$ and $b_M$ be random variables so that
$$
\lim_{M\rightarrow\infty} a_M \stackrel{\textrm{a.s.}}{=}a \mbox{ and } \lim_{M\rightarrow\infty} w_M \stackrel{\textrm{a.s.}}{=}b,
$$
where $a$ and $b$ are some constants. Then
\begin{equation}\label{eq:a alpha+b beta}
\lim_{M\rightarrow\infty} a_M \phi+b_M\beta\stackrel{\textrm{a.s.}}{=}a\phi+b\beta.
\end{equation}
Indeed, let $(\Omega, {\cal F}, P)$ be the probability space on which $a_M$ and $b_M$ are defined. Let further
\begin{align*}
{\cal A}&=\{\omega\in \Omega: \lim_{M\rightarrow \infty} a_M(\omega)=a\},\\
{\cal B}&=\{\omega\in \Omega: \lim_{M\rightarrow \infty} b_M(\omega)=b\},  \mbox{ and }\\
{\cal C}&=\{\omega\in \Omega: \lim_{M\rightarrow \infty} a_m(\omega)\phi+b_M(\omega)\beta=a\phi+b\beta\}.
\end{align*}
Then ${\cal A}\bigcap {\cal B}\subseteq {\cal C}$. So we have
\begin{align*}
\Pr({\cal C})\ge& \Pr({\cal A}\bigcap {\cal B})=\Pr(({\cal A}^c \bigcup {\cal B}^c)^c)\\
=& 1-\Pr({\cal A}^c\bigcup {\cal B}^c)\ge 1-\Pr({\cal A}^c)-\Pr({\cal B}^c)=1.
\end{align*}

  \IEEEproof (Theorem \ref{thm:PCP_D_ZF})
  The signal $y\kj$ defined in \eqref{eq:y^kj PCP0} can be written in the form:
  \begin{align} \label{eq:y^kj PCP}
&y^{[kj]}\nonumber \\
=&\sumlL {\sqrt{\rho_f}\over \lambda\kl_l}
\hat{\bg}_l^{[kl]^\dag}
 \bg_l\kj c_l^{[k]}
 +\sum_{{l=1}}^L \sum_{n=1\atop n\not =k}^K {\sqrt{\rho_f}\over \lambda\nl_l}  \hat{\bg}_l^{[nl]^\dag} \bg_l^{[kj]} c_l^{[n]} +w^{[kj]},
\end{align}
Let us denote the factors in front of $c_l^{[k]}$ in the first sum by
\begin{align*}
f_l\kj&={\sqrt{\rho_f}\over \lambda\kl_l}
\hat{\bg}_l^{[kl]^\dag}
 \bg_l\kj  \\
 &={\sqrt{\rho_f }\over \lambda\kl_l} \theta_{l}^{[kl]}
\left(\sum_{s=1}^L \sqrt{\rho_r\tau}\bg_l^{[ks]^\dagger}+\hat{\bw}_l^{[kl]^\dagger}\right)
\bg_{l}^{[kj]}.
 \end{align*}
 Opening the parenthesis and applying Lemma \ref{lem:LargeNumbers} to each term of the above expression, we obtain
 \begin{equation}\label{eq:f_l_kj}
 \lim_{M\rightarrow \infty} {1\over \sqrt{M}} f_l\kj\stackrel{\textrm{a.s.}}{=}{\sqrt{\rho_f\rho_r\tau}\over \eta_l^{[k]}}\beta\kj_l.
 \end{equation}
 Denote further the terms in the second sum of \eqref{eq:y^kj PCP} by
 \begin{align*}
 q_{nl}\kj&={\sqrt{\rho_f}\over \lambda\nl_l}  \hat{\bg}_l^{[nl]^\dag} \bg_l^{[kj]} c_l^{[n]}\\
 &={\sqrt{\rho_f }\over \lambda\kl_l} \theta_{l}^{[nl]}
\left(\sum_{s=1}^L \sqrt{\rho_r\tau}\bg_l^{[ns]^\dagger}+\hat{\bw}_l^{[nl]^\dagger}\right)
\bg_{l}^{[kj]}c_l^{[n]}
 \end{align*}
Again, using Lemma \ref{lem:LargeNumbers}, we obtain
\begin{equation}\label{eq:q_nl_kj}
 \lim_{M\rightarrow \infty} {1\over \sqrt{M}} q_{nl}\kj\stackrel{\textrm{a.s.}}{=}0.
 \end{equation}
 For $k=1,K$, denote
 $$
 F^{[k]}=\left(\begin{array}{ccc}
 f_1^{[k1]} &\ldots & f_L^{[k1]}\\
 \vdots & & \vdots\\
 f_1^{[kL]} & \ldots & f_L^{[kL]}
 \end{array} \right),
 $$
 and
 $$
 \by^{[k]}=
 \left( \begin{array}{l}
 y^{[k1]}\\
 \vdots \\
 y^{[kL]}
 \end{array}\right),
 \bc^{[k]}=\left( \begin{array}{l}
 c_1^{[k]}\\
 \vdots \\
 c_L^{[k]} \end{array}\right),\bs^{[k]}=\left( \begin{array}{l}
 s^{[k1]}\\
 \vdots \\
 s^{[kL]} \end{array}\right),
 $$
 and
 $$
 \bq^{[k]}= \left(\begin{array}{c}
 \sumlL \sum_{n=1 \atop n\not =k}^K q_{nl}^{[k1]}\\
 \vdots\\
 \sumlL \sum_{n=1 \atop n\not =k}^K q_{nl}^{[kL]}\end{array}\right),
 \bw^{[k]}=\left(\begin{array}{c}
 w^{[k1]}\\
 \vdots\\
 w^{[kL]}
 \end{array}\right).
 $$
 With this notations we have
 \begin{equation}\label{eq:y[k]}
 \by^{[k]}=F^{[k]}\bc^{[k]}+\bq^{[k]}+\bw^{[k]}=F^{[k]}\Phi^{[k]}\bs^{[k]}+\bq^{[k]}+\bw^{[k]}.
 \end{equation}
Let $V^{[k]}=F^{[k]}\Phi^{[k]}$. According to \eqref{eq:f_l_kj} we have
that the entries of $F^{[k]}$ almost surely converge to the corresponding entries of the matrix
$\sqrt{\rho_f\rho_r\tau}B_D^{[k]}$. From this, \eqref{eq:a alpha+b beta}, and from the definition of $\Phi^{[k]}$ it follows that the entries $v_{im}^{[k]}$ of $V^{[k]}$ have the property:
$$
\lim_{M\rightarrow\infty} {1\over \sqrt{M}} v_{im}^{[k]}\stackrel{\textrm{a.s.}}{=}\delta_{im}\sqrt{\rho_f\rho_r\rho_A\tau},\;i,m=1,L.
$$
From \eqref{eq:y[k]} and \eqref{eq:s\kj_M} we have
\begin{align*}
\hat{s}_M\kj=&{1\over \sqrt{M}} {1\over \sqrt{\rho_f\rho_r\rho_A\tau}}(v_{jj}^{[k]}s\kj+\sum_{l\not =j}^L v_{jl}^{[k]}s\kl\\
&+\sumlL \sum_{n=1\atop n\not =k}^K
q_{nl}\kj +w\kj).
\end{align*}
Applying now Lemma \ref{lem:inf_capacity} to the above expression we obtain
$$
\lim_{M\rightarrow \infty} I(\hat{s}_M\kj; s\kj)=H(s\kj).
$$
 \qed

 Thus, under our network assumptions defined in Section \ref{sec:PCP},  we constructed  a noise free
and interference free multi-cell LSAS with frequency reuse 1. In such LSAS the size of modulation ${\cal R}$ can be chosen arbitrary large and therefore the LSAS can achieve an arbitrary large capacity.

\subsection{Large-Scale Fading Decoding}\label{subsec:UPCP}
With the same Network Assumptions II we define the following protocol for uplink data transmission.

\noindent{\bf  Large-Scale Fading Decoding }
\begin{enumerate}
\item The $j$-th base station estimates the large scale fading coefficients $\beta_j^{[kn]},\;k=1,K,\;n=1,L$,
and sends them to the network controller.
\item The controller computes $L\times L$ decoding matrices
$$\Omega^{[k]}=\left(\begin{array}{c}
\underline{\omega}_1^{[k]}\\
\underline{\omega}_2^{[k]}\\
\vdots\\
\underline{\omega}_L^{[k]}
\end{array}\right),\; k=1,K,
$$
as functions of $\beta_j^{[kl]},\;j,l=1,L$.
\item The $j$-th base station computes  the MMSE estimates $\hat{\bg}_j^{[kj]},k=1,\ldots,K$
according to (\ref{eq:hat g}).
\item The $j$-th base station receives the vector $\by_j$ defined in \eqref{eq:yj} and computes the matched filtering estimates
 $$
 \tilde{x}^{[kj]}=\hat{\bg}_j^{[kj]^\dag}\by_j,\;k=1,\ldots,K,
 $$
 of the uplink signals $x^{[kj]}$
 (other options, for instance $M$-dimensional Zero Forcing or MMSE receiver, are possible here)
    and sends them to the network controller.
\item
The network controller computes the following estimates of $x\kj$
$$
\hat{x}^{[kj]}_M={1\over M \theta_j\kj\rho_r\sqrt{\tau}}
\underline{\omega}_j^{[k]}\left(\begin{array}{c}
\tilde{x}^{[k1]}\\
\vdots\\
\tilde{x}_{kL}
\end{array}\right),~k=1,\ldots,K.
$$
\end{enumerate}
\noindent{\bf The End}

We would like to point out the following things.
\begin{itemize}
\item Unlike the network MIMO approach, small-scale fading coefficients $h_{mj}^{[kj]}$ are used locally by the $j$-th base station and are not sent to the network controller.
\item Steps 1 and 2 are conducted only one time every large-scale coherence block, that is once every $T_\beta$ OFDM symbols.
\item
The estimates $\hat{\bg}_j^{[kl]}$ at Step 3 are computed one time for each coherence block, that is once every $T$ OFDM symbols.
\item
Step 4 and 5 are conducted for each channel use (each OFDM symbol).
\end{itemize}

Let
$$
\bB_\mathrm{U}^{[k]}=\left(\begin{array}{lll}
\beta_1^{[k1]} & \ldots & \beta_1^{[kL]}\\
\vdots & & \vdots \\
\beta_L^{[k1]} & \ldots & \beta_L^{[kL]}\end{array}
\right).
$$
Similar to the downlink case, we define {\em Zero-Forcing LSFD} as LSFD with
\begin{equation}\label{eq:U ZF-PCP}
 \Omega^{[k]}=\bB_\mathrm{U}^{[k]^{-1}},\;k=1,K.
\end{equation}

We again assume that signals $x\kj$ belong to a signal constellation ${\cal R}=\{r_1,\ldots,r_N\}$ with PMF $P_S(\cdot)$.

\begin{theorem}\label{thm:uplink ZF PCP} For Large-Scale Fading Decoding, we have
$$
\lim_{M\rightarrow \infty} I(\hat{x}^{[kj]}_M; x\kj)=H(x\kj).
$$

\end{theorem}
\IEEEproof
Using \eqref{eq:yj} and \eqref{eq:hat g}, we get
\begin{align*}
&\tilde{x}^{[kj]}=\hat{\bg}_j^{[kj]^\dagger} \by_j   \\
=&\sumlL x\kl \theta_j\kj \underbrace{\sumsL  \rho_r\sqrt{\tau} \bg_j\ksd\bg_j\kl}_{f_j\kl}\\
&+\underbrace{\theta_j\kj \sumsL (\sqrt{\rho_r\tau} \bg_j\ks+\hat{\bw}_j\kj)^\dagger (\sumlL
\sum_{n=1 \atop n\not =k}^K \sqrt{\rho_r}\bg_j\nl x\nl+\bw_j)}_{q\kj}
\end{align*}
 Applying Lemma 1 we obtain
\begin{align*}
\lim_{M\rightarrow \infty} {1\over M} f_j\kl & \stackrel{\textrm{a.s.}}{=}\theta_j\kj\rho_r\sqrt{\tau}\beta_j\kl, \mbox{ and } \\
\lim_{M\rightarrow \infty} {1\over M} q\kj & \stackrel{\textrm{a.s.}}{=}0.
\end{align*}
Denote
$$
 F^{[k]}=\left(\begin{array}{ccc}
 f_1^{[k1]} &\ldots & f_1^{[kL]}\\
 \vdots & & \vdots\\
 f_L^{[k1]} & \ldots & f_L^{[kL]}
 \end{array} \right),
 $$
 and
 $$
 \tilde{\x}^{[k]}=
 \left( \begin{array}{l}
 \tilde{x}^{[k1]}\\
 \vdots \\
 \tilde{x}^{[kL]}
 \end{array}\right),
 \x^{[k]}=\left( \begin{array}{l}
 x^{[k1]}\\
 \vdots \\
 x^{[kL]} \end{array}\right),\bq^{[k]}=\left( \begin{array}{l}
 q^{[k1]}\\
 \vdots \\
 q^{[kL]} \end{array}\right).
 $$
 We this notations we have
$$
 \tilde{\x}^{[k]}=F^{[k]}\x^{[k]}+\bq^{[k]}.
$$
Hence
\begin{align}
\left(
\begin{array}{c}
\hat{x}^{[k1]}_M\\
\vdots \\
\hat{x}^{[kL]}_M
\end{array}
\right)
=&{1\over M \theta_j\kj \rho_r\sqrt{\tau}}\Omega^{[k]}\tilde{\x}^{[k]}\nonumber \\
=&{1\over M \theta_j\kj \rho_r\sqrt{\tau}}(\Omega^{[k]}F^{[k]}\x^{[k]}+\Omega^{[k]}\bq^{[k]}).\label{eq:hatx_M^kj}
\end{align}
Note that as $M$ grows the matrix $F^{[k]}$ almost surely converges to the matrix $\theta_j\kj\rho_r\sqrt{\tau}B_U^{[k]}$.

Let $V^{[k]}=\Omega^{[k]}F^{[k]}$. Applying Lemma \ref{lem:LargeNumbers} and \eqref{eq:a alpha+b beta} to the entries of $V^{[k]}$, we get
$$
\lim_{M\rightarrow\infty} {1\over M \theta_j^{[kj]}\rho_r\sqrt{\tau}} v_{im}^{[k]}\stackrel{\textrm{a.s.}}{=}\delta_{im},\;i,m=1,L.
$$
From Lemma \ref{lem:LargeNumbers} and \eqref{eq:a alpha+b beta} it also follows that each entry of the vector $\Phi^{[k]}\bq^{[k]}$
almost surely converges to $0$.
Now applying Lemma \ref{lem:inf_capacity} to the entries of the vector \eqref{eq:hatx_M^kj} we obtain
$$
\lim_{M\rightarrow \infty} I(\hat{x}_M\kj; x\kj)=H(x\kj).
$$
 \qed

Again we obtained a noise free
and interference free multi-cell LSAS, that is an LSAS with arbitrary large capacity,  with frequency reuse 1.

\subsection{Estimation of Large-Scale Fading Coefficients}\label{subsec:Estimation of Large-Scale Coef}
In this subsection we outline one possible algorithm for estimation of large-scale fading coefficients.

First, we would like to remind that the coefficients $\beta_j^{[kl]}$ do not depend
on antenna indices as well as on OFDM tone indices. Thus, between any given base station and a mobile, there is only one large-scale fading coefficient. Second, these coefficients
change only when a mobile significantly change its geographical location. The standard assumption is that in the radius of
$10$ wavelengths, the large-scale fading coefficients are approximately constant. In contrast, small-scale fading coefficients significantly change as soon as a user moves by
a quarter of the wavelength.
Thus, large-scale fading coefficients change about $40$ times slower than small-scale fading coefficients.

Let us enumerate all users across the entire network by integers from $1$ to $LK$. Denote by $\beta_j^{[i]}$ and
$\bh_j^{[i]}$ the large-scale fading coefficient and fast fading vector between the $i$-th users and the $j$-th base station.
Let further $\bv^{[r]},~r=1,\mu$, be a set of mutually orthogonal $\mu$-tuples of norm $1$.
Since the coefficient $\beta_j^{[i]}$ does not depend on OFMD tone indices,
 it is enough if the $i$-th user transmits a training sequence, say $\bv^{[1]}$, in only one OFDM tone. We assume that no other user transmits $\bv^{[1]}$ in this OFDM tones and that users that transmit $\bv^{[r]}, r\ge 2$,  in this
 OFDM tone are enumerated by $2, \ldots,\mu$.
In this OFDM tone the $j$-th base station receives the signal
$$
\bY_j=\sqrt{\rho_r\mu}\sqrt{\beta_{j}^{[i]}}\bh_{j}^{[i]} \bv^{[i]^\dag}+\sum_{k=2}^\mu
\sqrt{\rho_r \mu}\sqrt{\beta_j^{[k]}}\bh_j^{[k]}\bv^{[k]}+ \bW_j,
$$
where $\bW_j\in \mathbb{C}^{M\times \mu}$ is the additive white Gaussian noise matrix with i.i.d. ${\cal CN}(0,1)$
 entries.
In order to estimate $\beta_j^{[i]}$, the $j$-th base station  first computes
$$
\by=\bY_j\bv^{[1]}=\sqrt{\rho_r\mu}\sqrt{\beta_j\kl}\bh_j\kl+\hat{\bw}_j,
$$
and further
$$
\hat{\beta}_j^{[i]}={1\over M\rho_r\mu}\by^\dag \by-{1\over \rho_r\mu}.
$$
Taking into account that $\hat{\bw}_j\sim {\cal CN}(0,\mbf I_M)$ and using Lemma \ref{lem:LargeNumbers}, after simple computations, we obtain
$$
\lim_{M\to \infty} \hat{\beta}_j\kl \stackrel{\textrm{a.s.}}{=}\beta_j\kl.
$$
Training sequences transmitted in different OFDM tones do not interfere with each other.
Thus if $N$ is the number of OFDM tones the above approach allows us to have $N\mu$ non interfering training sequences. For example if $N=1400,\mu=8$, and $K=20$  this approach
allows one to estimate large-scale fading coefficients in a network of $L=560$ cells.
In real life application we need to estimate large-scale coefficients only of the users located in neighboring cells and users located far away from each other can reuse the same OFDM tone and training sequence.


\section{Finite $M$ Analysis}\label{sec:finiteM}
In this section we derive SINR expressions for downlink and uplink LSFPs in which the number $M$ of base station antennas appears as a parameter. We do not assume ZF-LSFP, instead we consider generic LSFPs in which matrices $\Phi^{[k]}$ and $\Omega^{[k]}$ can have arbitrary entries, with the constraints that the transmit powers of each base station and each user, in average, do not exceed $\rho_f$ and $\rho_r$ respectively.

First, we would like to note that though the $j$-th base stations needs only MMSE estimates $\hat{\bg}_j\kj,k=1,K$,
it can also compute the MMSE estimates of the channel vectors $\hat{\bg}_j\kl,\;k=1,K,\;l=1,L$, between itself and all the users in the network as
\begin{equation}\label{eq:hat gjkl}
\hat{\bg}_j^{[kl]}=\bY_j(\theta_j^{[kl]}\br^{[k]})=\theta_{j}^{[kl]}\sqrt{\rho_r\tau}
\sum_{s=1}^L \bg_j^{[ks]}+\hat{\bw}\kl_j,
\end{equation}
where
$$
\theta_j^{[kl]}={\sqrt{\rho_r\tau}\beta_j^{[kl]}\over 1+\rho_r\tau\sum_{s=1}^L \beta_j^{[ks]}},
$$
and $\hat{\bw}\kl_j\sim {\cal CN}(0,\theta_j^{[kl]^2} \mbf I_M)$. Let
$$
\tilde{\bg}_j\kl=\bg-\hat{\bg}_j\kl
$$
be the estimation error.

The following properties of $\hat{\bg}_j^{[kl]}$ and $\tilde{\bg}_j\kl$ are either well known (see for example \cite{Kay}) or
can be easily derived from their definitions.
\begin{enumerate}
\item  If $(j,k)\not =(j',k')$ then the vectors $\hat{\bg}_j\kl$ and $\hat{\bg}_{j'}^{[k's]}$ are independent for any $l$ and $s$.
\item The vectors $\hat{\bg}_j\kl$ and $\tilde{\bg}_i^{[ns]}$ are uncorrelated for any indices $j,k,l,i,n$, and $s$.
\item
The vectors $\hat{\bg}_j\kl$ and $\tilde{\bg}_j\kl$ have the following distributions:
\begin{align}
\hat{\bg}_j\kl&\sim {\cal CN}\left(0,{\rho_r\tau\beta_j^{[kl]^2}\over 1+\sumsL \rho_r\tau\beta_j\ks} \mbf I_M\right),\label{eq:hatg_jkl^2} \\
&\mbox{ and } \nonumber \\
 \tilde{\bg}_j\kl&\sim {\cal CN}\left(0,\left(\beta_j\kl- {\rho_r\tau\beta_j\kl\over 1+\sumsL \rho_r\tau\beta_j\ks}\right) \mbf I_M\right).
 \label{eq:tildeg_jkl^2}
 \end{align}
\item It is not difficult to show that
\begin{equation}\label{eq:EjklEjkj}
\mathbb{E}[ \hat{\bg}_j\kl \hat{\bg}_j\kjd]={\rho_r\tau\beta_j\kl\beta_j\kj\over 1+\sumsL \rho_r\tau\beta_j\ks}\mbf I_M.
\end{equation}
\end{enumerate}

\subsection{Performance of Large-Scale Fading Precoding with Finite $M$}\label{subsec:LSFP with fininite M}

Denote the $(j,v)$ entry of matrix $\Phi^{[k]}$ by $\phi_j\kv$.
It will be convenient for us to assume that in LSFP protocol the normalization coefficients ${1\over \lambda\kj_j}$, used in Step \ref{step:DL PCP small-scale} of LSFP,
are absorbed into the $\phi_j^{[kv]}$.
In other words, we replace $\phi_j^{[kv]}$ with $\phi_j^{[kv]}\over \lambda\kj_j$ and in Step \ref{step:DL PCP small-scale} use $\bu_j^{[k]}=\hat{\bg}_j\kjd$. This allows us to shorten notations.

According to the  LSFP protocol, the $k$-th terminal in the $l$-th cell receives the signal
\begin{align}\label{eq:ykl}
y\kl=&\sqrt{\rho_f}\sum_{j=1}^L \x_j{\bg}_j\kl+w\kl\nonumber\\
=&\sum_{j=1}^L \sum_{n=1}^K \hat{\bg}_j^{[nj]^\dag}c_j^{[n]}{\bg}_j\kl+w\kl.
\end{align}

Taking into account that
\begin{align}
\bg_j\kl&=\hat{\bg}_j\kl+\tilde{\bg}_j\kl, \mbox{ and } \nonumber \\
c_j^{[k]}&=\sumvL \phi_j\kv s\kv,\label{eq: c_j^k=sum ...}
\end{align}
 and replacing the random variable in front of
$s\kl$ with its expected value we obtain:
\begin{align*}
&y\kl\nonumber\\
=&\sqrt{\rho_f}\sum_{j=1}^L \sum_{n=1}^K \hat{\bg}_j^{[nj]^\dagger}\hat{\bg}_j\kl c_j^{[n]} +
\sqrt{\rho_f}\sum_{j=1}^L  \sum_{n=1}^K \hat{\bg}_j^{[nj]^\dagger}\tilde{\bg}_j\kl c_j^{[n]}\nonumber\\
&+w\kl
\end{align*}
\begin{align*}
=& s\kl \sqrt{\rho_f} \sumjL \phi_j\kl \hat{\bg}_j\kjd\hat{\bg}_j\kl\nonumber\\
&+\sqrt{\rho_f}\sum_{v=1\atop v\not=l}^L s\kv\sumjL \phi_j\kv \hat{\bg}_j\kjd\hat{\bg}_j\kl\nonumber\\
&+\sqrt{\rho_f}\sum_{j=1}^L\sum_{{n=1\atop n\not =k}}^{K}
\hat{\bg}_j^{[nj]^\dagger}\hat{\bg}_j\kl c_j^{[n]}+\sqrt{\rho_f}\sum_{j=1}^L  \sum_{n=1}^K \hat{\bg}_j^{[nj]^\dagger}\tilde{\bg}_j\kl c_j^{[n]}\nonumber\\
&+w\kl\nonumber\\
\end{align*}
\begin{align}\label{eq:ykl1}
=& \underbrace{s\kl  \sqrt{\rho_f}\sumjL \phi_j\kl \mathbb{E}[\hat{\bg}_j\kjd\hat{\bg}_j\kl]}_{T_0}\nonumber\\
&+\underbrace{\sqrt{\rho_f} s\kl \sumjL \phi_j\kl (\hat{\bg}_j\kjd\hat{\bg}_j\kl-\mathbb{E}[\hat{\bg}_j\kjd\hat{\bg}_j\kl])}_{T_1}\nonumber\\
&+
\underbrace{\sqrt{\rho_f} \sum_{v=1\atop v\not=l} s\kv\sumjL \phi_j\kv \hat{\bg}_j\kjd\hat{\bg}_j\kl}_{T_2}\nonumber\\
&+\underbrace{\sqrt{\rho_f} \sum_{j=1}^L\sum_{{n=1\atop n\not =k}}^{K}
\hat{\bg}_j^{[nj]^\dagger}\hat{\bg}_j\kl c_j^{[n]}}_{T_3}\nonumber\\
&+\underbrace{\sqrt{\rho_f} \sum_{j=1}^L  \sum_{n=1}^K \hat{\bg}_j^{[nj]^\dagger}\tilde{\bg}_j\kl c_j^{[n]}}_{T_4}
+\underbrace{w\kl}_{T_5}.
\end{align}
To make notations shorter, we denote the five terms (four sums and $w\kl$)
 in above expression by $T_0, \ldots, T_5$ respectively.

First, we note that these terms are mutually uncorrelated.
Indeed, since $s\kl$ is independent of any $\hat{\bg}_i^{[ns]}$, we have
\begin{align*}
&\mathbb{E}[T_0^\dag T_1]\\
=&\mathbb{E}[s\kld s\kl]
\rho_f\sum_{j=1}^L \sum_{i=1}^L
\phi_{j}\kld \phi_{i}\kl \mathbb{E}[\hat{\bg}_{j}\kjd\hat{\bg}_{j}\kl]^\dag \\
&\cdot
\mathbb{E}[(\hat{\bg}_{i}\kid\hat{\bg}_{i}\kl-\mathbb{E}[\hat{\bg}_{i}\kid\hat{\bg}_{i}\kl])]
=0.
\end{align*}
Since $s\kl$ is independent of $s^{[nv]}$ if $(k,l)\not = (n,v)$ we get
\begin{align*}
&\mathbb{E}[T_0^\dag T_2]=0, \mathbb{E}[T_0^\dag T_3]=0, \mathbb{E}[T_1^\dag T_2]=0, \\
& \mathbb{E}[T_1^\dag T_3]=0, \mathbb{E}[T_2^\dag T_3]=0.
\end{align*}
Since $\tilde{\bg}_j\kl$ is uncorrelated with any $\hat{\bg}_i^{[nv]}$, it is not difficult to check
that $T_4$ is uncorrelated with $T_0,T_1,T_2$, and $T_3$. Finally, $T_5=w\kl$ is independent
from all other terms.
Thus, we can rewrite (\ref{eq:ykl1}) as
$$
y\kl=s\kl \sqrt{\rho_f}\sumjL \phi_j\kl \mathbb{E}[\hat{\bg}_j\kjd\hat{\bg}_j\kl]     + w\kl_{eff},
$$
where the effective noise has the variance
$$
\mbox{Var}[w\kl_{eff}]=\mbox{Var}[T_1]+ \mbox{Var}[T_2] +\mbox{Var}[T_3]+ \mbox{Var}[T_4]+ \mbox{Var}[T_5].
$$
Using (\ref{eq:EjklEjkj}), we obtain
\begin{align}\label{eq:numerator}
&\sqrt{\rho_f} \sumjL \phi_j\kl \mathbb{E}[\hat{\bg}_j\kjd\hat{\bg}_j\kl] \nonumber\\
=&\sqrt{\rho_f} M \rho_r\tau \sum_{j=1}^L {\beta_j\kl\beta_j\kj\over
1+\sum_{s=1}^L \rho_r \tau\beta_j\ks} \phi_j\kl,
\end{align}
which defines the quantity $\epsilon\kl_l$ in Step \ref{step:DL PCP epsilon} of the LSFP protocol.

According to the protocol the $l$-th base station sends $\epsilon\kl_l$ to the corresponding user. So the user can use $\epsilon\kl_l$ to detect signal $s\kl$.
  Note that $\epsilon\kl_l$  depends
only on the statistical parameters of the channel and not on  instantaneous channel realizations. According to \cite[Theorem 1]{hassibi}, the worst-case uncorrelated additive
noise is independent Gaussian noise with the same variance. Hence the downlink rate $R\kl$ can be lower bounded as follows
{\small \begin{align}\label{eq:Rkl_1st_expr}
&R\kl=I(y\kl;s\kl\left| \epsilon\kl_l \right.)
\nonumber\\
\ge& \log_2\left(1+ { |\epsilon\kl_l|^2 \over
\mbox{Var}[T_1]+ \mbox{Var}[T_2] +\mbox{Var}[T_3]+ \mbox{Var}[T_4]+ \mbox{Var}[T_5]}\right).
\end{align}}

Now we proceed with finding the variances $\mbox{Var}[T_j],j=1,5$.

 The term $T_1$ is caused by the {\bf channel uncertainty}.  The $k$-th user located in the $l$-th cell
 does not know
the actual value of the effective channel $\hat{\bg}_j\kjd\hat{\bg}_j\kl$, but only the expected value $\mathbb{E}[\hat{\bg}_j\kjd\hat{\bg}_j\kl]$. So the difference, uncertainty,
between the actual and expected values of the effective channel contributes to the interference.   To estimate the variance of $T_1$ we first note that
 the signals $s\kl$ and $s\nv$ are independent if $(k,l)\not=(n,v)$. Next, we note that
for any $1\le k,n \le K$, and $1\le l,v \le L$, the vectors $\hat{\bg}_j\kl$ and $\hat{\bg}_{i}\nv$ are independent
if $j\not=i$. Taking this into account we obtain
\begin{align*}
&\mbox{Var}[T_1]\nonumber\\
=&\mbox{E}[s\kl s\kld]\sumjL \sumiL \phi_j\kl\phi_i\kld
\mathbb{E}\left[ (\hat{\bg}_j\kjd\hat{\bg}_j\kl-\mathbb{E}[\hat{\bg}_j\kjd\hat{\bg}_j\kl])\right.\nonumber\\
&\left.\cdot(\hat{\bg}_i\kid\hat{\bg}_i\kl-\mathbb{E}[\hat{\bg}_i\kid\hat{\bg}_i\kl])^\dag\right]\nonumber\\
=&\sumjL |\phi_j\kl|^2 \mbox{Var}[ \hat{\bg}_j\kjd\hat{\bg}_j\kl].
\end{align*}
For computing
$\mbox{Var}[\hat{\bg}_j^{[kj]^\dagger}\hat{\bg}_j\kl]$
we note  that according to (\ref{eq:hat gjkl}) $\hat{\bg}_j^{[kj]}$ is proportional to
$\hat{\bg}_j\kl$, namely
\begin{equation}\label{eq:gjkj is propor gjkl}
\hat{\bg}_j^{[kj]}={\theta_j\kj\over \theta_j\kl} \hat{\bg}_j\kl.
\end{equation}
Let $\bz=(z_1,\ldots,z_M)^\top\sim {\cal CN}(0,\mbf I_M)$.
It is well know that
$$
z_i\sim {\cal CN}(0,1) \mbox{ and } z_i^\dag z_i \sim \Gamma(3,1),
$$
and therefore $\mbox{Var}[z_i^\dag z_i]=1$.
Using this, (\ref{eq:gjkj is propor gjkl}), and (\ref{eq:hatg_jkl^2}), we get
\begin{align}\label{eq:Var}
&\mbox{Var}\left[\hat{\bg}_j^{[kj]^\dagger}\hat{\bg}_j^{[kl]}\right]\nonumber \\
=&
{\rho_r\tau\beta_j\klt\over 1+\sumsL \rho_r\tau\beta_j\ks}
{\rho_r\tau\beta_j\kjt\over 1+\sumsL \rho_r\tau\beta_j\ks}
\mbox{Var}[\bz^\dagger \bz]\nonumber \\
=&{\rho_r\tau\beta_j\klt\over 1+\sumsL \rho_r\tau\beta_j\ks}
{\rho_r\tau\beta_j\kjt\over 1+\sumsL \rho_r\tau\beta_j\ks}\cdot M.
\end{align}
Thus
\begin{align}
\label{eq:VT1}
&\mbox{Var}[T_1]\nonumber\\
=&M\sumjL \left|\phi_j\kl\right|^2 {\rho_r\tau\beta_j\klt\over 1+\sumsL \rho_r\tau\beta_j\ks}
{\rho_r\tau\beta_j\kjt\over 1+\sumsL \rho_r\tau\beta_j\ks}.
\end{align}

Next, we consider the term $T_2$. This term is caused by the {\bf pilot contamination effect}. Since the $k$-th users
in cells $j$ and $l$ use the same training sequence
the vector $\hat{\bg}_j\kj$ is correlated with the vector $\hat{\bg}_j\kl$ even if $l\not =j$.
Taking into
account the same facts about signals $s\nv$ and vectors $\hat{\bg}_i\nv$ that we used in the derivation of $\mbox{Var}[T_1]$, we obtain
\begin{align*}
&\mbox{Var}[T_2]\\
=&\sum_{v=1\atop v\not = l}^l \mbox{E}[s\kv s\kvd] \sumjL \sumiL \phi_j\kv\phi_i\kvd\mathbb{E}
[\hat{\bg}_j\kjd\hat{\bg}_j\kl \hat{\bg}_i\kld\hat{\bg}_i\ki]\nonumber\\
=&\sum_{v=1\atop v\not = l}^l  \sumjL \sum_{i=1\atop i\not =j}^L \phi_j\kv\phi_i\kvd \mathbb{E}[\hat{\bg}_j\kjd\hat{\bg}_j\kl]
\mathbb{E}[\hat{\bg}_i\kld\hat{\bg}_i\ki]\\
&+\sum_{v=1\atop v\not = l}^l  \sumjL  |\phi_j\kv|^2 \mathbb{E}[\hat{\bg}_j\kjd\hat{\bg}_j\kl \hat{\bg}_j\kld\hat{\bg}_j\kj].
\end{align*}

Using (\ref{eq:Var}) and (\ref{eq:EjklEjkj}), we obtain
\begin{align*}
&\mathbb{E}[\hat{\bg}_j\kjd\hat{\bg}_j\kl \hat{\bg}_j\kld\hat{\bg}_j\kj]\\
=&\mbox{Var}[\hat{\bg}_j\kjd\hat{\bg}_j\kl]+|\mathbb{E}[\hat{\bg}_j\kjd\hat{\bg}_j\kl]|^2\\
=&M \cdot {\rho_r\tau\beta_j\klt\over 1+\sumsL \rho_r\tau\beta_j\ks}\cdot
{\rho_r\tau\beta_j\kjt\over 1+\sumsL \rho_r\tau\beta_j\ks}\\
 &+\left({M\rho_r\tau\beta_j\kl\beta_j\kj\over 1+\sumsL \rho_r\tau\beta_j\ks}\right)^2.
\end{align*}

From this, we have
\begin{align}\label{eq:VarT2}
&\mbox{Var}[T_2]\nonumber\\
=&\sum_{v=1\atop v\not=l}^L \left|\sumjL {M \rho_r\tau\beta_j\kj \beta_j\kl \over 1+\sumsL \rho_r\tau\beta_j\ks} \phi_j\kv\right|^2\nonumber\\
& + \sum_{v=1\atop v\not =l}^L M \sumjL {\rho_r\tau\beta_j\klt\over 1+\sumsL \rho_r\tau\beta_j\ks}
{\rho_r\tau\beta_j\kjt\over 1+\sumsL \rho_r\tau\beta_j\ks}|\phi_j\kv|^2.
\end{align}

Let us consider now the term $T_3$, which is caused by the {\bf nonorthogonality of channel vectors}.
 In the asymptotic regime, as $M$ tends to infinity, the normalized inner-product of vectors $\hat{\bg}_j\kl$ and $\hat{\bg}_j^{[nj]}$ almost surely converges to zero. For finite $M$, however, this is not the case and $T_3$  may significantly contribute to the interference. From \eqref{eq: c_j^k=sum ...}, we have
$$
\mathbb{E}[|c_j^{[n]}|^2]=\sumvL \sum_{u=1}^L \phi_j\nv\phi_j^{[nu]^\dag} \mathbb{E}
[s\nv s^{[nu]^\dag}]= \sumvL |\phi_j\nv|^2.
$$
Using this, (\ref{eq:EjklEjkj}), and the fact that $\hat{\bg}\nj_j$ and $\bg\kj_j$ are uncorrelated,   we obtain
\begin{align*}
&\mbox{Var}[T_3]\\
=&\sum_{j=1}^L\sum_{{n=1\atop n\not =k}}^{K}
\mathbb{E}[\hat{\bg}_j^{[nj]^\dag}\hat{\bg}_j^{[kl]}\hat{\bg}_j^{[kl]^\dag}\hat{\bg}_j^{[nj]}]
\cdot \mathbb{E}[|c_j^{[n]}|^2]\\
=&\sum_{j=1}^L\sum_{{n=1\atop n\not =k}}^{K}  \mathbb{E}[\mbox{Tr}(\hat{\bg}_j^{[nj]}\hat{\bg}_j^{[nj]^\dag}\hat{\bg}_j^{[kl]}\hat{\bg}_j^{[kl]^\dag})]
\cdot \mathbb{E}[|c_j^{[n]}|^2]\\
=&\sum_{j=1}^L\sum_{{n=1\atop n\not =k}}^{K} \mbox{Tr}\left(\mathbb{E}[\hat{\bg}_j^{[nj]}\hat{\bg}_j^{[nj]^\dag}]\mathbb{E}[\hat{\bg}_j^{[kl]}\hat{\bg}_j^{[kl]^\dag})]\right)
\cdot \mathbb{E}[|c_j^{[n]}|^2]\\
=&  M \sum_{j=1}^L\sum_{{n=1\atop n\not =k}}^{K}
{\rho_r\tau\beta_j^{[kl]^2}\over 1+\sum_{s=1}^L \rho_r\tau \beta_j^{[ks]}}
\cdot
{\rho_r\tau\beta_j^{[nj]^2}\over 1+\sum_{s=1}^L \rho_r\tau \beta_j^{[ks]}}\\
&\cdot\sum_{v=1}^L |\phi_j\kv|^2.
\end{align*}

Term $T_4$ is caused by  {\bf estimation errors of channel vectors}. Since
$\hat{\bg}_j\kl$ is uncorrelated with any $\tilde{\bg}_i\nv$, using (\ref{eq:hat gjkl}) and (\ref{eq:tildeg_jkl^2}) we obtain
\begin{align*}
&\mbox{Var}[T_4]= \sum_{j=1}^L  \sum_{n=1}^K  \mbox{Tr}\left(\mathbb{E}[\hat{\bg}_j^{[nj]}\hat{\bg}_j^{[nj]^\dag}]\mathbb{E}[
\tilde{\bg}_j^{[kl]} \tilde{\bg}_j^{[kl]^\dag} ]\right)\cdot \mathbb{E}[|c_j^{[n]}|^2]\\
=&M \sum_{j=1}^L  \sum_{n=1}^K
\left(\beta_j^{[kl]}-{\rho_r\tau \beta_j^{[kl]^2}\over 1+\sum_{s=1}^L \rho_r\tau \beta_j^{[ks]}}\right)\\
&\cdot
{\rho_r\tau\beta_j^{[nj]^2}\over 1+\sum_{s=1}^L \rho_r\tau\beta_j\ks} \cdot \sum_{v=1}^L |\phi_j\kv|^2.
\end{align*}

Finally, the {\bf power of additive noise} $w\kl$ is
$
\mbox{Var}[T_5]=\mbox{Var}[w\kl]=1.
$

Now, after some computations, we obtain from (\ref{eq:Rkl_1st_expr}) the following theorem.
\begin{theorem}\label{thm:downlink SINRkl} If the conjugate beamforming precoding is used in Step \ref{step:DL PCP small-scale} of LSFP then
the downlink transmission rate $R\kl$ is lower bounded by
$$
R\kl_D \ge \log_2(1+\mbox{SINR}\kl_D),\;k=1,K,\;l=1,L,
$$
where
\begin{equation}\label{eq:downlink_SINRkl}
\mbox{SINR}\kl_D={\rho_f M^2 \rho_r^2\tau^2 \left|\sum_{j=1}^L {\beta_j\kl\beta_j\kj\over
1+\sum_{s=1}^L \rho_r \tau\beta_j\ks} \phi_j\kl\right|^2 \over
M^2 I_1 + M I_2 +1},
\end{equation}
and
$$
I_1=\rho_f\rho_r^2\tau^2 \sum_{v=1\atop v\not=l}^L \left|\sum_{j=1}^L {\beta_j\kl\beta_j\kj\over
1+\sum_{s=1}^L \rho_r\tau \beta_j\ks }\phi_j\kv\right|^2
$$
and
$$
I_2=\rho_f\rho_r\tau\sum_{j=1}^L \sum_{n=1}^K {\beta_j^{[nj]^2}
\over 1+\sum_{s=1}^l\rho_r\tau
\beta_j^{[ns]}}\beta_j\kl\sum_{v=1}^l|\phi_j^{[nv]}|^2.
$$
\end{theorem}

It is instructive to apply Theorem \ref{thm:downlink SINRkl} for the cases
when LSFP is not used and
when ZF-LSFP is used.

\subsection{No LSFP and ZF-LSFP}\label{subsec:NoLSFP and ZF-LSFP}

If we do not use LSFP then the matrices $\Phi^{[k]}$ are diagonal. Hence, taking into account (\ref{eq:EjklEjkj})
and the power constraint
$$
\mathbb{E}[||\hat{\bg}_j\kj c_j^{[k]}||^2]=\mathbb{E}[||\hat{\bg}_j\kj \phi_j\kj s\kj||^2]= 1,
$$
we conclude that
\begin{equation}\label{eq:noPCPalpha}
\phi_j\kl=\delta_{jl}{(1+\sum_{s=1}^L\rho_r\tau\beta_j^{[ns]})^{1/2}\over \sqrt{M\rho_r\tau}\beta_j\kj},
\end{equation}
where $\delta_{jl}$ is Kronecker's delta.
 The numerator of \eqref{eq:downlink_SINRkl} will have the form
$$
\rho_f M\rho_r \tau {\beta\klt_l\over 1+\sumsL \rho_r\tau \beta_l\ks}
$$
and
\begin{align*}
I_1=&{\rho_f\rho_r\tau\over M}\sum_{v=1\atop v\not =l}^L {\beta\klt_v\over 1+\sumsL \rho_r\tau \beta\ks_v},\\
I_2=&{\rho_f\over M} \sum_{j=1}^L\sum_{n=1}^K {\beta\njt_j\over 1+\sumsL \rho_r \tau \beta\ns_j}
\beta\kl_j\\
&\sumvL {1+\sumsL \rho_r\tau \beta\ns_j\over \rho_r\tau
\beta\nj_j}.
\end{align*}

Let now  $M\rightarrow \infty$. In this case we get
$$
\lim_{M\rightarrow \infty }
\mbox{SINR}\kl_{D,NO~LSFP}={\beta\kl_l/(1+\sumsL \rho_r \tau\beta\ks_l)\over
\sum_{v\not =l}^L \beta\klt_v/(1+\sumsL \rho_r\tau\beta\ks_v)}.
$$
Thus we again obtained (\ref{eq:SINR1}) and again we see that
for very large number of base station antennas the interference is completely defined by the pilot contamination effect.

Let us consider ZF-LSFP. Let
$$
\mu_j^{[k]}={\beta\kj_j\over 1+\sumsL \rho_r\tau \beta\ks_j}
$$
and define $L\times L$ matrix
\begin{equation}\label{eq:newBD}
\bB^{[k]}=\left(\begin{array}{lll}
\beta_{1}^{[k1]}\mu_1^{[k]}& \ldots & \beta_{L}^{[k1]}\mu_L^{[k]}\\
\vdots & & \vdots \\
\beta_{1}^{[kL]}\mu_1^{[k]} & \ldots & \beta_{L}^{[kL]}\mu_L^{[k]}
\end{array}
\right).
\end{equation}
Let further
$$
\Phi^{[k]}=\sqrt{\rho_A}\bB^{[k]^{-1}},k=1,K,
$$
where $\rho_A$ is a normalization factor to insure the constraint \eqref{eq:||alpha_j||<1}.
With such $\Phi^{[k]}$ the numerator of \eqref{eq:downlink_SINRkl} and $I_1$ are
$$
\rho_f M^2\rho_r^2\rho_A\tau^2 \mbox{ and }I_1=0.
$$
In the asymptotic regime, as $M\rightarrow \infty$ we get
$$
\lim_{M\rightarrow \infty }
\mbox{SINR}\kl_{D,ZF-LSFP}=\lim_{M\rightarrow \infty } {\rho_f M^2\rho_r^2\tau^2\over M I_2}=\infty.
$$
So, similar to Theorem \ref{thm:PCP_D_ZF}, we obtained that in the asymptotic regime ZF-LSFP allows achieving infinite SINRs for all users.

\begin{remark}
 The matrix \eqref{eq:newBD} has a slightly different form compared with \eqref{eq:BDk} because we assumed that the user knows only the expected gain of the effective channel, that is the quantity $\epsilon\kl_l$. In contrast, in Section \ref{sec:PCP} it is implicitly assumed that the user knows the actual value of the effective channels $\hat{\bg}\kl_j\bg\kl_j$.
\end{remark}

\subsection{Optimization Problem}\label{subsec:optimization problem}
We can slightly simplify notation if we replace $\phi_j\kl$ with
$$
\alpha_j^{[kl]}={\sqrt{\rho_r\tau}\beta_j\kj\over 1+\sum_{s=1}^L \rho_r \tau \beta_j\ks} \phi_j\kl.
$$
This replacement does not cause any problems since
all the quantities used in the above expression are assumed being known at the $j$-th
base station. After this replacement we get
\begin{equation}\label{eq:downlinkSINRkl_1}
\mbox{SINR}\kl_D={M\rho_f\rho_r\tau \left|\sum_{j=1}^L \beta_j\kl\alpha_j^{[kl]}\right|^2 \over
M J_1 + J_2 +1/M},
\end{equation}
where
$$
J_1=\rho_f\rho_r\tau \sum_{v=1\atop v\not=l}^L \left|\sum_{j=1}^L \beta_j\kl  \alpha_j^{[kv]}\right|^2,
$$
and
$$
J_2=\rho_f\sum_{j=1}^L \sum_{n=1}^K \beta_j\kl(1+\sum_{s=1}^L
\rho_r\tau \beta_j^{[ns]})\left(\sum_{v=1}^L \left|\phi_j^{[nv]^*}\right|^2\right).
$$
With this notations the average power of the $j$-th base station is the
following function of the large-scale fading coefficients and the LSFP coefficients:
\begin{align}\label{eq:Power of BS-j}
\gamma_j=&\sum_{k=1}^K
\mathbb{E}
[|\hat{\bg}_j\kj c_j^{[k]}|^2]\nonumber\\
 =& M \sum_{k=1}^K(1+\sum_{s=1}^L\rho_r\tau \beta_j^{[ks]})
(\sum_{v=1}^L |\alpha_j^{[kv]}|^2).
\end{align}
Using \eqref{eq:Power of BS-j} one may formulate different optimization problems with base station power constraints.
In particular, in Part II of this paper \cite{Part II} we consider the following problem
$$
\max_{\alpha_j^{[nr]},n=1,K,\;j,r=1,L}\; \min_{k=1,K,\;l=1,L} \mbox{SINR}\kl_D
$$
subject to the constraints
$$
\gamma_j\le 1,\;j=1,L.
$$

\subsection{First Simulation Results}\label{subsec:first simul results}
\begin{figure}[htb]
\centering
\includegraphics[scale=0.35]{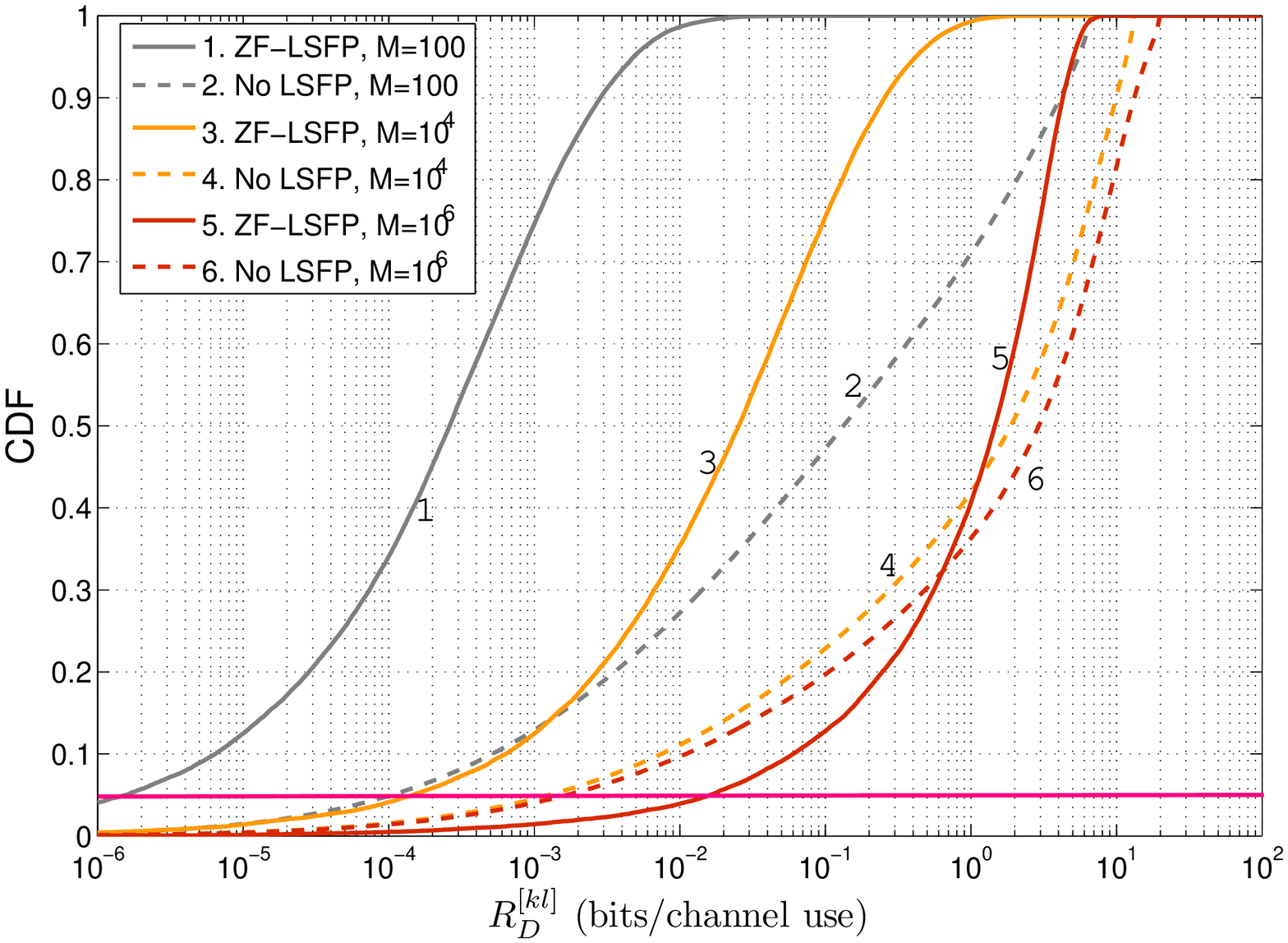}
\caption{The CDF of the achievable rate $R_D^{[kl]}=\log\left(1+\mbox{SINR}^{[kl]}\right)$ for two existing methods with different number of antennas.}
\label{fig-rate}
\end{figure}

\begin{figure}[htb]
\centering
\includegraphics[scale=0.35]{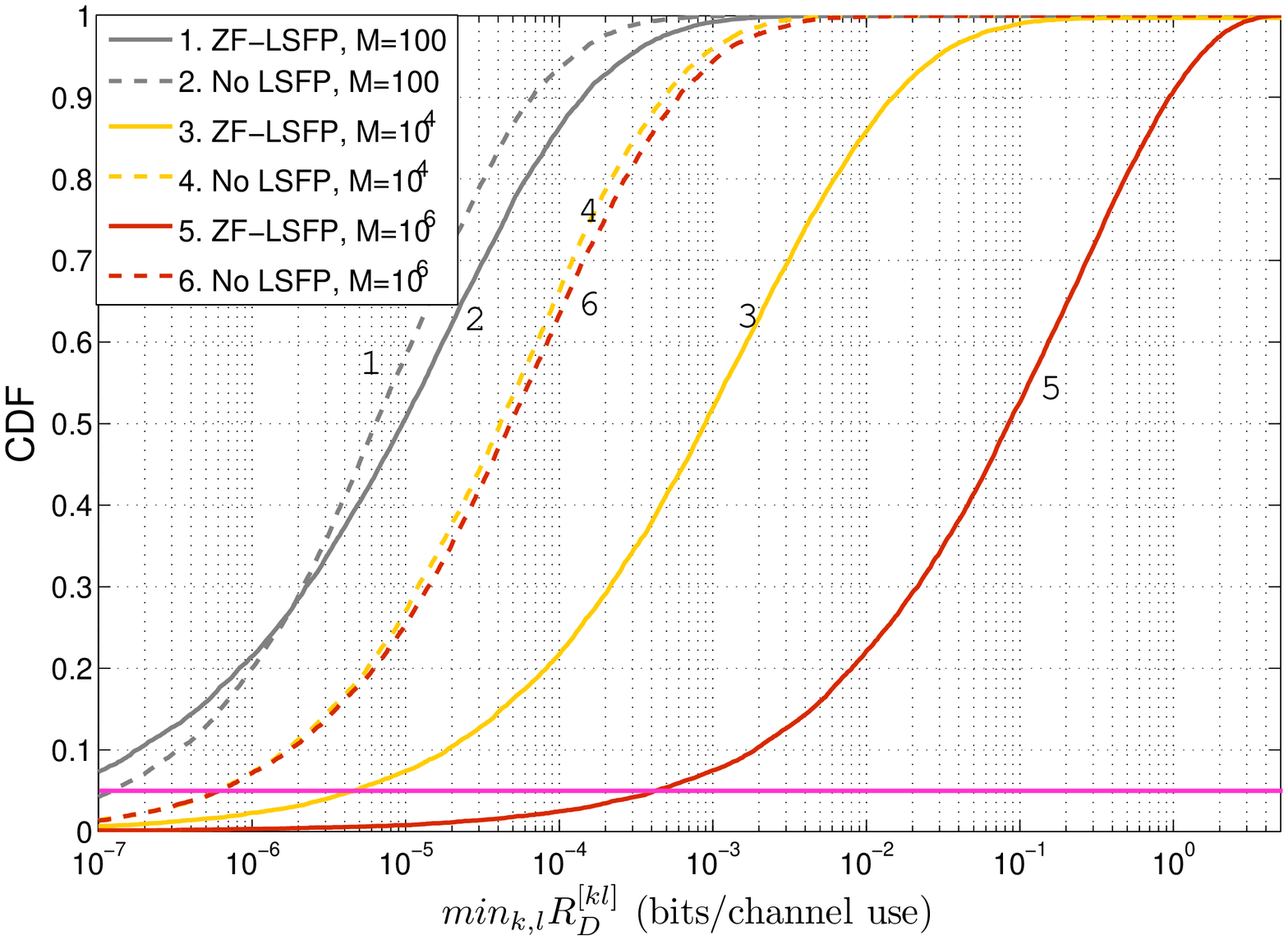}
\caption{The CDF of the achievable rate $R_D^{[kl]}=\log\left(1+\mbox{SINR}^{[kl]}\right)$ for two existing methods with different number of antennas.}
\label{fig-minrate}
\end{figure}

Since ZF-LSFP allows achieving infinite SINRs when $M\rightarrow \infty$ it is natural  to ask what it gives us when $M$ is finite. In order to answer this question  we generate random large-scale fading coefficients and use Theorem \ref{eq:downlink_SINRkl}
for finding corresponding downlink rates $R\kl_D$ when LSFP is not used (No LSFP) and when ZF-LSFP is used.
In Fig.~\ref{fig-rate} and Fig.~\ref{fig-minrate}, we present simulation results for the CDFs of $R\kl_D, k=1,K,\;l=1,L$,  and the CDF of the minimum rate $\min_{k,l} R\kl_D$, respectively. We plot achievable rates and CDF in horizontal and vertical axis, respectively. In these simulations, we assumed $K=10$.

  It can be observed from both figures that by increasing the number of antennas we significantly improve the performance of ZF-LSFP. In Fig.~\ref{fig-rate}, ZF-LSFP achieves $5\%$-outage rate around $ 10^{-6}$ bits per channel use at $M=100$;  the achievable rate is improved to around $10^{-2}$ bits per channel use with $M=10^6$ antennas. On the other hand, the achievable rate of no-LSFP is saturated when $M$ is getting very large.

  According to Fig.~\ref{fig-rate}, when we consider the rates of all users,  the $5\%$-outage rate of no-LSFP is larger than the $5\%$-outage rate of ZF-LSFP at $M=100,10^4$ and only at $M=10^5$ ZF-LSFP starts outperforming No LSFP. In the case of CDFs of minimum rates $\min_{k,l} R\kl_D$ shown in Fig. \ref{fig-minrate}, ZF-LSFP has better performance than No LSFP already at $M=10^4$, which is still a very large number of antennas.

  Thus we conclude  that in a regime with a practical number of antennas, e.g. $M=100$ and smaller, the ZF-LSFP performs worse than No LSFP approach. It is a natural question to ask whether an LSFP different from ZF-LSFP can
  give any gain in such scenario, or LSFP is only a theoretical tool for investigation of the asymptotic case $M\rightarrow \infty$. We answer to this question in Part II of the paper \cite{Part II}. We will show that properly designed
  LSFP gives very significant gain over other methods of interference reduction.


\subsection{Performance of Large-Scale Fading Decoding with finite $M$}\label{subsec:LSFD with fininte M}
Denote the $(l,v)$-th entry of matrix $\Omega^{[k]}$ by $\omega_l^{[kv]}$.
Similar to the downlink case we obtain the following theorem.
\begin{theorem} \label{thm:UL SINR}
$$
R_{U}\kl\ge \log_2(1+SINR_{UL}\kl),\;k=1,K,\;l=1,L,
$$
where
\begin{equation}\label{eq:U_SINR}
\mbox{SINR}_{U}\kl={M^2\rho_r^3\tau^2 \left|\sumvL {\beta_v\kv\beta_v\kl\over 1+\sumsL\rho_r\tau\beta_v\ks}
\omega_l\kv\right|^2\over M^2 I_1+M I_2},
\end{equation}
and
\begin{align*}
I_1=&\rho_r^3\tau^2\sum_{j=1\atop j\not=l}^L \left|\sumvL {\beta_v\kv\beta_v\kj\over 1+\sumsL\rho_r\tau\beta_v\ks}
\omega_l\kv\right|^2, \\
I_2=&\sumvL |\omega_l\kv|^2 {\rho_r\tau \beta_v^{[kv]^2}\over
1+\sumsL \rho_r\tau\beta_v\ks} \left(1+\rho_r \sumjL  \sumnK \beta_v\nj\right).
\end{align*}
\end{theorem}
Before presenting a formal proof of this theorem we would like to note that
the important distinction of this result from Theorem \ref{eq:downlink_SINRkl}. The coefficients $\omega_l\kv$ appear   only in $\mbox{SINR}_{U}\kl$
and not in any other $\mbox{SINR}_{U}^{[nj]}$. Thus optimal LSFP coefficients $\omega_l\kv,v=1,L$, can be chosen independently for
each user, which significantly simplify search of optimal coefficients.

We also would like to note that we can slightly simplify expression for $\mbox{SINR}_{U}\kl$ if replace $\omega_l\kv$ with
$$
\omega_l^{[kv]^*}={\sqrt{\rho_r\tau M} \beta_v\kv \over 1+\sumsL \rho_r \tau \beta_v\ks}\omega_l\kv.
$$
In this case we get
\begin{equation}\label{eq:UL_SINR_simplified}
\mbox{SINR}_{U}\kl= {M\rho_r^2\tau \left| \sumvL \beta_v\kl \omega_l^{[kv]^*}\right|^2 \over M J_1 + J_2},
\end{equation}
where
$$
J_1=\rho_r^2\tau \sum_{j=1\atop j\not =l}^L \left|\sumvL \beta_v\kj \omega_l^{[kv]^*}\right|^2
$$
and
$$
J_2=\sumvL |\omega_l^{[kv]^*}|^2 (1+\sumsL \rho_r\tau \beta_v\ks) (1+\rho_r\sumjL \sum_{n=1}^K \beta_v\nj).
$$

Finally, it is instructive to consider the expression (\ref{eq:UL_SINR_simplified}) for the cases of NO~LSFP and ZF-LSFP.
If we do not use LSFP then $\omega_l^{[kv]}=\delta_{lv}$. Substituting such $\omega_l^{[kv]}=\delta_{lv}$ into (\ref{eq:UL_SINR_simplified}),
in the regime $M\rightarrow \infty$ we get for   the expression (\ref{eq:uplinkSINR}) from Section \ref{subsec:Pilot Contamination}. In the case of  ZF-LSFP we chose $\omega_l\kv$ according
to (\ref{eq:U ZF-PCP}).  Using these coefficients in (\ref{eq:UL_SINR_simplified}) we get the result of Theorem \ref{thm:uplink ZF PCP}  from Section \ref{sec:PCP}.

The proof is similar to the proof of Theorem \ref{thm:downlink SINRkl}. However, as we noted above,
the LSFP coefficients $\omega_l\kv$ appear in $\mbox{SINR}_{U}\kl$ in a different way than in Theorem
\ref{eq:downlink_SINRkl}. So it is important to carefully track all indices involved into the computations.
The proof is in Appendix B.

\section{Appendix A}
Let ${\cal R}=\{r_1,\ldots,r_N\}$ be a constellation of signals such that $\min_{i,j} \mbox{dist}(r_i,r_j)\ge \Delta$ for some positive real $\Delta$ and $N$ is an arbitrary integer.

Before presenting a proof of Lemma \ref{lem:inf_capacity}, we would like to note that when we deal with mutual information or entropy functions we not always can  replace a random variable, say $x_M$, with its limit value. For example, let $x_M$ be defined by
\begin{align*}
&\Pr(x_M=-1-1/M)=\Pr(x_M=-1+1/M)\\
=&\Pr(x_M=1-1/M)=\Pr(x_M=1+1/M)=1/4.
\end{align*}
It is easy to see that
$$
\lim_{M\rightarrow\infty} x_M\stackrel{\textrm{a.s.}}{=}x,
$$
where $x$ is a random variable defined by
$$\Pr(x=-1)=\Pr(x=1)={1\over 2}.
$$
At the same time, in the case of the binary entropy, we have
$$H_2(x)=-\sum_{x\in \{-1,1\}} \Pr(x)\log_2 \Pr(x)=1,
$$
 while
$H_2(x_M)=2$ for any $M$ and therefore
$$
\lim_{M\rightarrow\infty} H_2(x_M)=2.
$$
Thus we have
$$
H_2(\lim_{M\rightarrow\infty} x_M)\not =\lim_{M\rightarrow\infty} H_2(x_M).
$$
This example shows that the statement of Lemma \ref{lem:inf_capacity} is not trivial and indeed
 needs a proof.

Let us define for some positive $\phi$ and $\beta$  the events $A_M$, $W_M$ by $a_M \in [1-\phi,1+\phi]$ and $w_M\in [-\beta,\beta]$ respectively.
We  would like to remind that
$$
\lim_{M\rightarrow \infty} a_M \stackrel{\textrm{a.s.}}{=}1 \mbox{ and }
\lim_{M\rightarrow \infty} w_M \stackrel{\textrm{a.s.}}{=}0
$$
imply  that
\begin{align}
\lim_{M\rightarrow \infty} \Pr(A_M^c)&=0, \mbox{ and }  \label{eq:Pr(AMc)=0}\\
\lim_{M\rightarrow \infty} \Pr(W_M^c)&=0. \label{eq:Pr(WMc)=0}
\end{align}
We also have
\begin{equation}\label{eq:Pr(AM and BM)=1}
\lim_{M\rightarrow \infty} \Pr(A_M \bigcap W_M) =1.
\end{equation}
Indeed
\begin{align*}\label{eq:AM intersect WM=1}
&\Pr(A_M \bigcap W_M)=\Pr((A_M^c \bigcup W_M^c)^c)\nonumber\\
=&1-\Pr(A_M^c \bigcup W_M^c)\ge 1-\Pr(A_M^c)
-\Pr(W_M^c)=1.
\end{align*}

In a similar way one can show that
\begin{equation}\label{eq:Pr(AM and BMc)=0}
\lim_{M\rightarrow \infty} \Pr(A_M \bigcap  W_M^c) =0.
\end{equation}

\IEEEproof (Lemma \ref{lem:inf_capacity})   Below we show that for any given $\Delta$, by choosing sufficiently large $M$, we can make $I(s_M;s)$ being arbitrary close to $H(s)$.

To make notation short we assume that
 $N$ is even and the signals ${\cal R}$ are evenly spaced on the real line that is
$$
{\cal R}=\{-N/2\cdot \Delta, -(N/2-1)\Delta, \ldots, (N/2-1)\Delta,N/2\cdot\Delta\}.
$$
Generalizations for unevenly spaced signals and for complex signals are straightforward.

For a given $s_M$,  let $q_M\in {\cal R}$ be such that
$$
\mbox{dist}(q_M,s_M)\le \mbox{dist}(s',s_M) \mbox{ for any }s'\in {\cal R}\setminus q_M.
$$
In other words $q_M$ is obtained by demodulation of $s_M$ with respect to ${\cal R}$. According to the data processing inequality we have
\begin{equation}\label{eq:Iq<Is}
I(q_M,s)\le I(s_M,s) \le H(s).
\end{equation}
Next,
$$
I(q_M,s)=H(s)-H(s|q_M),
$$
Denote by $P_Q(\cdot)$ the PMF of $q_M$. Then
\begin{align*}
&H(s|q_M)\\
=& \sum_{q_M\in {\cal R}} P_Q(q_M) \sum_{s\in {\cal R}} P_{S|Q}(s|q_M)\log P_{S|Q}(s|q_M)\\
=&\sum_{q_M\in {\cal R}} P_Q(q_M) [P_{S|Q}(s=q_M|q_M)\log P_{S|Q}(s=q_M|q_M)\\
 &  +\sum_{s'\in {\cal R}\setminus q_M} P_{S|Q}(s'|q_M)\log P_{S|Q}(s'|q_M)]
\end{align*}
The conditional probability $P_{S|Q}(s=q_M|q_M)$ can be written as
\begin{align*}
&P_{S|Q}(s=q_M|q_M)\\
=&{P_S(q_M)P_{Q|S}(q_M|s=q_M) \over P_S(q_M)P_{Q|S}(q_M|s=q_M)+\sum_{s'\in {\cal R}\setminus q_M} P_S(s')
P_{Q|S}(q_M|s')}.
\end{align*}
First, we estimate $P_S(q_M)P_{Q|S}(q_M|s'),s'\not =q_M$. From
the definition of $q_M$ it follows that
$$
  P_{Q|S}(q_M|s')=\Pr(a_Ms'+w_M\in (q_M-\Delta/2,q_M+\Delta/2)).
  $$
Let us assume that
$q_M=(n-1)\Delta$  and $s'=n\Delta$ for some positive integer $n$.
If $a_M\ge 1-1/4n$. Then in order to have
  \begin{align*}
  a_Ms'+w_M \in& [q_M-\Delta/2,q_M+\Delta/2]\\
  &=[(n-1)\Delta-\Delta/2,(n-1)\Delta+\Delta/2],
\end{align*}
we need that $w_M<-\Delta/4$. Similarly, if $a_M\le 1+1/4n$ we have to have
$w_M>-7\Delta/4$. Hence, if $a_M\in [1-1/4n,1+1/4n]$ then $w_M$ can not be outside
of the interval $(-7\Delta/4 , -\Delta/4  )$. Thus
\begin{align*}
&P_{Q|S}(q_M=(n-1)\Delta|s'=n\Delta)\\
\le &\Pr(a_M\in [1-1/4n,1+1/4n] \mbox{ and } w_M\in (-7/4\Delta,-1/4\Delta))\\
&+\Pr(a_M\not \in [1-1/4n,1+1/4n]).
\end{align*}
Applying now \eqref{eq:Pr(AM and BMc)=0} and \eqref{eq:Pr(AMc)=0} to the above terms, we get
$$
\lim_{M\rightarrow\infty} P_{Q|S}(q_M=(n-1)\Delta|s'=n\Delta)=0.
$$
For any other $s'\in {\cal R}\setminus q_M$ we can use similar arguments that lead to
$$
\lim_{M\rightarrow\infty} P_{Q|S}(q_M|s')=0.
$$
Hence, taking into account that the number of terms in the sum
is countable, we get
\begin{align}\label{eq:sum s' not = qM}
\lim_{M\rightarrow\infty} \sum_{s'\in {\cal R}\setminus q_M} P_S(s')P_{Q|S}(q_M|s')=0.
\end{align}

Using the same type of arguments we can show that
\begin{align*}
&P_{Q|S}(q_M|s=q_M)\\
=&\Pr(a_Mq_M+w_M\in (q_M-\Delta/2,q_M+\Delta/2))\\
\ge& \Pr(a_M\in [1-1/4n,1+1/4n] \mbox{ and } w_M\in [-\Delta/4,\Delta/4]).
\end{align*}
Applying \eqref{eq:Pr(AM and BM)=1} to this lower bound we conclude that it converges to $1$. Thus
\begin{equation}\label{eq:s = qM}
 \lim_{M\rightarrow\infty} P_{Q|S}(q_M|s=q_M)=1.
\end{equation}
From \eqref{eq:s = qM}
 and \eqref{eq:sum s' not = qM} we get
 $$
 \lim_{M\rightarrow\infty} P_{S|Q}(s=q_M|q_M)=1.
$$
and further
\begin{equation}\label{eq:P log P=0,1}
 \lim_{M\rightarrow\infty} P_{S|Q}(s=q_M|q_M)\log P_{S|Q}(s=q_M|q_M)=0.
\end{equation}

For $s'\not =q_M$ we have
\begin{align*}
&P_{S|Q}(s'|q_M)\\
=&{P_S(s')P_{Q|S}(q_M|s') \over P_S(q_M)P_{Q|S}(q_M|s=q_M)+\sum_{s''\in {\cal R}\setminus q_M} P_S(s'')
P_{Q|S}(q_M|s'')}.
\end{align*}
Using \eqref{eq:sum s' not = qM} and \eqref{eq:s = qM} we obtain
 $$
\lim_{M\rightarrow\infty} P_{S|Q}(s'|q_M)=0,
$$
and further
\begin{equation}\label{eq:P log P=0,2}
 \lim_{M\rightarrow\infty} P_{S|Q}(s'|q_M)\log P_{S|Q}(s'|q_M)=0.
\end{equation}
From \eqref{eq:P log P=0,1} and \eqref{eq:P log P=0,2} it follows that
$$
\lim_{M\rightarrow\infty} H(s|q_M)=0 \mbox{ and } \lim_{M\rightarrow\infty} I(q_M;s)=H(s),
$$
which, together with \eqref{eq:Iq<Is}, finishes the proof.
\qed

\section{Appendix B}
\IEEEproof of Theorem \ref{thm:UL SINR}.
The $l$-th base station receives the signal
$$
\by_l=\sqrt{\rho_r}\sum_{j=1}^L \sum_{n=1}^K \bg_l\nj x\nj+\bw_l.
$$
After applying the matched filter $\hat{\bg_l}\kl$ it gets
$$
\tilde{x}\kl=\hat{\bg}_l^{[kl]^\dag}\by_l=\sqrt{\rho_r}\sum_{j=1}^L \sum_{n=1}^K
\hat{\bg}_l^{[kl]^\dag}\bg_l\nj x\nj+\hat{\bg}_l^{[kl]^\dag}\bw_l,
$$
where $\bw_l\sim {\cal CN}(0,\mbf I_M)$.
The network controller collects these estimates and applies pilot contamination decoding:
\begin{align*}
&\hat{x}\kl=\sum_{v=1}^L \omega_l\kv \tilde{x}\kv\nonumber\\
 =& \sqrt{\rho_r}\sumvL\sumjL \sumnK \omega_l\kv \hat{\bg}_v^{[kv]^\dag}
\bg_v\nj x\nj+\sumvL \omega_l\kv \hat{\bg}_v^{[kv]^\dag}\bw_v\\
=& \sqrt{\rho_r}\sumvL\sumjL \sumnK \omega_l\kv \hat{\bg}_v^{[kv]^\dag}
\hat{\bg}_v\nj x\nj\nonumber \\
&+\sqrt{\rho_r}\sumvL\sumjL \sumnK \omega_l\kv \hat{\bg}_v^{[kv]^\dag}
\tilde{\bg}_v\nj x\nj+\sumvL \omega_l\kv \hat{\bg}_v^{[kv]^\dag}\bw_v\nonumber\\
=& s\kl \sqrt{\rho_r} \sumvL \omega_l\kv \hat{\bg}_v\kvd\hat{\bg}_v\kl\nonumber \\
&+\sqrt{\rho_r} \sumvL\sum_{j=1\atop j\not= l} \sumnK
\omega_l\kv \hat{\bg}_v\kv\hat{\bg}_v\nj s\nj\nonumber\\
&+ \sqrt{\rho_r} \sumvL \sumjL \sum_{n=1 \atop n\not =k} \omega_l\kv \hat{\bg}_v\kv\hat{\bg}_v\nj s\nj\nonumber\\
&+\sqrt{\rho_r}\sumvL\sumjL \sumnK \omega_l\kv \hat{\bg}_v^{[kv]^\dag}
\tilde{\bg}_v\nj x\nj+\sumvL \omega_l\kv \hat{\bg}_v^{[kv]^\dag}\bw_v\nonumber\\
\end{align*}
\begin{align}
=& s\kl \sqrt{\rho_r} \sumvL \omega_l\kv \mathbb{E}[\hat{\bg}_v\kvd\hat{\bg}_v\kl]\nonumber \\
 &+ s\kl \sqrt{\rho_r} \sumvL \omega_l\kv (\hat{\bg}_v\kvd\hat{\bg}_v\kl-\mathbb{E}[\hat{\bg}_v\kvd\hat{\bg}_v\kl])\nonumber \\
&+\sqrt{\rho_r} \sumvL\sum_{j=1\atop j\not= l} \sumnK
\omega_l\kv \hat{\bg}_v\kvt\hat{\bg}_v\nj s\nj\nonumber\\
&+ \sqrt{\rho_r} \sumvL \sumjL \sum_{n=1 \atop n\not =k} \omega_l\kv \hat{\bg}_v\kvt\hat{\bg}_v\nj s\nj\nonumber\\
&+\sqrt{\rho_r}\sumvL\sumjL \sumnK \omega_l\kv \hat{\bg}_v^{[kv]^\dag}
\tilde{\bg}_v\nj x\nj+\sumvL \omega_l\kv \hat{\bg}_v^{[kv]^\dag}\bw_v.\label{eq:UP_main_expres}
\end{align}

Denote the terms of this expression by $Q_0, \ldots, Q_5$. Similar to the downlink case it is not
difficult to prove that these terms are mutually uncorrelated. Hence we can rewrite (\ref{eq:UP_main_expres}) in the form:
$$
\hat{x}\kl=s\kl \sqrt{\rho_r} \sumvL \omega_l\kv \mathbb{E}[\hat{\bg}_v\kvd\hat{\bg}_v\kl]+w\kl_{eff},
$$
and
\begin{align*}
&\mbox{Var}[w\kl_{eff}]\\
=&\mbox{Var}[Q_1]+\mbox{Var}[Q_2]+\mbox{Var}[Q_3]+\mbox{Var}[Q_4]+\mbox{Var}[Q_5].
\end{align*}
Again using \cite[Theorem 1]{hassibi} we obtain the following lower bound on the uplink rate $R\kl_U$
\begin{align*}
&R\kl_U=I\left(\hat{x}\kl; \by_l
\left| \sqrt{\rho_r} \sumvL \omega_l\kv \mathbb{E}[\hat{\bg}_v\kvd\hat{\bg}_v\kl]\right. \right)\\
\ge & \log_2\left(1+{\rho_r \left|\sumvL \omega_l\kv \mathbb{E}[\hat{\bg}_v\kvd\hat{\bg}_v\kl]\right|^2\over
\mbox{Var}[Q_1]+\mbox{Var}[Q_2]+\mbox{Var}[Q_3]+\mbox{Var}[Q_4]+\mbox{Var}[Q_5]}\right)
\end{align*}
Below we find the variances of $Q_1, \ldots, Q_5$.

The term $Q_1$ is caused by the uncertainty of $l$-the base station about the affective channel $\hat{\bg}_l\kl \hat{\bg}_l\kl$.
Since $\hat{\bg}_v^{[ks]}$ and $\hat{\bg}_r^{[ni]}$ are uncorrelated for any $k,s,n,i$, using (\ref{eq:Var}), we get
\begin{align*}
&\mbox{Var}[Q_1]=\rho_r \mbox{E}[|s\kl|^2]\sumvL \mbox{Var}[\hat{\bg}_v\kv\hat{\bg}_v\kl]\\
=&\rho_r M\sumvL |\omega_l\kl|^2 {\rho_r\tau\beta_v\kvt\over 1+\sumsL \rho_r \tau\beta_v\ks}
\cdot {\rho_r\tau\beta_v\klt\over 1+\sumsL \rho_r \tau\beta_v\ks}.
\end{align*}

Term $Q_2$ is caused by the pilot contamination. In order to find
$\mbox{Var}[Q_2]$ we note that $s\nj$ and $s^{[mv]}$ are independent if $(n,j)\not=(m,v)$ and that
$\hat{\bg}_v\nj$ and $\hat{\bg}_v^{[mv]}$ are uncorrelated if $n\not =m$. Now, using  (\ref{eq:EjklEjkj}) and (\ref{eq:Var}), we obtain
\begin{align*}
&\mbox{Var}[Q_2]\\
=&\rho_r \sumvL \sum_{j=1 \atop j\not=l}^L \sum_{r=1\atop r\not=v}^L \omega_l\kv\omega_l^{[kr]^\dag}
\mathbb{E}[\hat{\bg}_v\kvt \hat{\bg}_v\kv]\mathbb{E}[\hat{\bg}_r\kjt \hat{\bg}_r\kj]    \\
&+\rho_r \sumvL \sum_{j=1 \atop j\not=l}^L |\omega_l\kv|^2 \mathbb{E}[\hat{\bg}_v\kvt \hat{\bg}_v\kj \hat{\bg}_v\kjt \hat{\bg}_v\kv]\\
=&\rho_r  \sum_{j=1 \atop j\not=l}^L \left|\sumvL {M\rho_r\tau \beta_v\kv\beta_v\kj\over 1+\sumsL \rho_r\tau\beta_v\ks}\omega_l\kv\right|^2\\
&+ \rho_r M \sum_{j=1 \atop j\not=l}^L\sumvL {\rho_r\tau \beta_v\kjt\over 1+\sumsL \rho_r\tau \beta_v\ks}\cdot
{\rho_r\tau \beta_v\kvt\over 1+\sumsL \rho_r\tau \beta_v\ks}\\
&\cdot |\omega_l\kv|^2
\end{align*}

Consider term $Q_3$. In the asymptotic regime, as $M$ tends to infinity, the normalized inner-product of $\hat{\bg}_v\kv$ and $\hat{\bg}_v\nj$
almost surely tends to zero since for $n\not =k$ these vectors are independent. For finite $M$, however, we can not ignore the interference caused by $Q_3$.
To compute $Q_3$ we use the same fact as we used for $Q_2$ and, using (\ref{eq:EjklEjkj}), obtain
\begin{align*}
&\mbox{Var}[Q_3]\\
=&\rho_r\sumvL \sumjL \sum_{n=1 \atop n\not =k}^K  |\omega_l\kv|^2 \mbox{Tr}\left(
\mathbb{E}[\hat{\bg}_v\kv\hat{\bg}_v\kvd]\mathbb{E}[\hat{\bg}_v\nj\hat{\bg}_v\njt]\right)\\
&\cdot\mathbb{E}[|s\nj|^2]\\
=&\rho_r M \sumvL \sumjL \sum_{n=1 \atop n\not =k}^K |\omega_l\kv|^2
{\rho_r\tau \beta_v\kvt\over 1+\sumsL \rho_r\tau \beta_v\ks}\\
&\cdot {\rho_r\tau \beta_v\njt\over 1+\sumsL \rho_r\tau \beta_v\ks}.
\end{align*}

Term $Q_4$ is caused by the estimation error $\tilde{\bg}_v\nj$. Taking into account
that in the case of MMSE estimation the estimate $\hat{\bg}_v\nj$ and estimation error $\tilde{\bg}_v\nj$
are uncorrelated, and using (\ref{eq:EjklEjkj}) and (\ref{eq:tildeg_jkl^2}), we obtain
\begin{align*}
&\mbox{Var}[Q_4]\\
=&\rho_r\sumvL \sumjL \sum_{n=1}^K |\omega_l\kv|^2 \mbox{Tr}\left(\mathbb{E}[\hat{\bg}_v\kv \hat{\bg}_v\kvt]
\mathbb{E}[\tilde{\bg}_v\nj \hat{\bg}_v\njt]\right)\\
&\cdot \mathbb{E}[|s\nj|^2]\\
=&\rho_r M \sumvL \sumjL \sum_{n=1}^K |\omega_l\kv|^2 {\rho_r\tau \beta_v\kvt\over 1+\sumsL \rho_r\tau\beta_v\ks}\\
&\cdot\left(\beta_v\nj-{\rho_r\tau\beta_v\njt\over 1+\sumsL \rho_r\tau \beta_v^{[ns]}}\right).
\end{align*}

Finally, term $Q_5$ is caused by the additive noise at the receivers of base stations.
Using the same arguments as above, we obtain
\begin{align*}
\mbox{Var}[Q_5]=&
\sumvL |\omega_l\kv|^2 \mbox{Tr}\left(\mathbb{E}[\hat{\bg}_v\kv \hat{\bg}_v\kvt]\mathbb{E}[\bw_v\bw_v^\dag]\right) \\
=&M\sumvL |\omega_l\kv|^2  {\rho_r\tau \beta_v\kvt\over 1+\sumsL \rho_r\tau\beta_v\ks}.
\end{align*}
Combining the obtained results finishes the proof. \qed

\vspace{0.2cm}
\noindent{\bf Acknowledgement} The authors would like to thank Carl Nuzman for his help with the proof of Lemma \ref{lem:inf_capacity}.


\begin{thebibliography}{99}
\bibitem{Marzetta10} T.~L.~Marzetta, "Multi-cellular wireless with base stations employing
unlimited numbers of antennas," in {\em Proc. UCSD Inf. Theory Applicat.
Workshop,} Feb. 2010.

\bibitem{Part II} L.~Li, A.~Ashikhmin, T.~Marzetta, ``Interference Reduction in Massive MIMO Systems II: Downlink Analysis for a Finite Number of Antennas,'' submitted to {\em IEEE Trans. on Information Theory}.

\bibitem{Marzetta11}
T.~L.~Marzetta, ``Noncooperative Cellular Wireless with Unlimited Numbers of Base
  Station Antennas,'' \emph{, IEEE Trans. on Wireless Communications},
  {\bf 9},  pp. 3590 --3600, 2010.

\bibitem{Ngo} H.~Q.~Ngo, E.~G.~Larsson, and T.~L.~Marzetta,``Energy and spectral efficiency
of very largemultiuserMIMOsystems,'' {\em IEEE Trans. on Commununications,}
{\bf 61}, pp.~1436-–1449, 2013.

\bibitem{Xu} G.~Y.~Li, Z.-K.~Xu, C.~Xiong, C.-Y.~Yang, S.-Q.~Zhang, Y.~Chen, and
S.-G.Xu,``Energy-efficient wireless communications: Tutorial, survey,
and open issues,'' {\em IEEE Wireless Commun. Mag.,} {\bf 18}, pp.~28–-35, 2011.

\bibitem{Xion} C.~Xiong, G.~Y.~Li, S.~Zhang, Y.~Chen, and S.~Xu, ``Energy- and spectral-
efficiency tradeoff in downlink OFDMA networks,'' {\em IEEE Trans.
Wireless Commun.,} {\bf 10}, pp.~3874-–3886, 2011.

\bibitem{Rusek} F.~Rusek, D.~Persson, B.~K.~Lau, E.~G.~Larsson, T.~L.~Marzetta, O.~Edfors,
and F.~Tufvesson, ``Scaling up MIMO: Opportunities and challenges
with very large arrays,'' {\em IEEE Signal Process. Mag.,} {\bf 30}, pp.~40-–46, 2013.

\bibitem{Larsson}  E.~G.~Larsson, F.~Tufvesson, O.~Edfors, and T.~L.~Marzetta, ``Massive
MIMO for next generation wireless systems,'' {\em IEEE Commun. Mag.,}
{\bf 52}, pp.~186-–195, 2014.

\bibitem{Lu} Lu~Lu ; G.~Y.~Li, A.~L.~Swindlehurst, A.~Ashikhmin, Rui~Zhang,
``An Overview of Massive MIMO: Benefits and Challenges,''
{\em IEEE Journal of Selected Topics in Signal Processing,}
{\bf 8}, pp.~742--758, 2014.



\bibitem{Fernandec} F.~Fernandes, A.~Ashikhmin, T.~L.~Marzetta, ``Interfernce Reduction on Cellular Networks with Large Antenna Arrays,''
{\it IEEE International Conference on Communicaions (ICC),} Ottawa, Canada, 2012.

\bibitem{Fernandec1} F.~Fernandes, A.~Ashikhmin, T.~L.~Marzetta,
 ``Inter-Cell Interference in Noncooperative TDD Large Scale Antenna Systems,''
{\it IEEE Journal on Selected Areas in Communications,} {\bf 31}, pp.~192--201, 2013.

\bibitem{Appaih} K.~Appaiah, A.~Ashikhmin, T.~L.~Marzetta,
``Included in Your Digital Subscription Pilot Contamination Reduction in Multi-User TDD Systems,'' {\em International Conference on Communications}, pp. 1--5, 2010.

\bibitem{hoydis} J.~Hoydis, S.~ten~Brink, M.~Debbah, ``Massive-MIMO: How Many Antennas do We Need,'' {\em arXiv:1107.1709v2}.

\bibitem{huh} H.~Huh, G.~Caire, H.C.~Papadopoulos, S.A.~Ramprashad, ``Achieving ``Massive-MIMO'' Spectral
Efficiency with a Not-so-Large Number of Antennas,'' {\em arXiv:1107.3862v2}.





\bibitem{patent} A.~Ashikhmin and T.~Marzetta, ``Large-Scale Antenna Method And Apparatus Of Wireless Communication With Suppression Of Intercell Interference,'' {\em US8774146 patent}, filed on Dec. 19, 2011, Issued on July 8, 2014.

\bibitem{Ashikhmin2012} A.~Ashikhmin and T.~L.~Marzetta, ``Pilot contamination precoding in multi-cell large scale antenna systems,'' {\em Proceedings of 2012 IEEE International Symposium on Information Theory  (ISIT),}
 pp.~1137--1141, 2012.


\bibitem{Hong2013} H.~Yang,~T.~L.~Marzetta, ``Total energy efficiency of cellular large scale antenna system multiple access mobile networks,'' {\em IEEE Online Conference on Green Communications (GreenCom),} pp.27--32, 2013.   

\bibitem{Kay} S.~M.~Kay, {\em Fundamentals of Statistical Sygnal Processing. I: Estimation Theory}, Prentice Hall PTR, 1993.

\bibitem{Jose}  J.~Jose,  A.~Ashikhmin, T.~L.~Marzetta, S.~Vishwanath,
``Pilot Contamination and Precoding in Multi-Cell TDD Systems,''
{\em IEEE Transactions on Wireless Communications,}
vol. 10, pp. 2640 -- 2651, 2011.

\bibitem{NetworkMIMO1} M.~K.~Karakayali, G.~J.~Foschini, R.~A.~Valenzuela, and R.~D.~Yates,
"On the maximum common rate achievable in a coordinated network,"
{\em IEEE International Conf. of Commun.,} vol. 9, pp. 4333 -– 4338, Jun.
2006.

\bibitem{NetworkMIMO2} M.~K.~Karakayali, G.~J.~Foschini, R.~A.~Valenzuela,  "Network coordination
for spectrally efficient communications in cellular systems,"
{\em IEEE Trans. Wireless Commun.,} vol. 13, no. 4, pp. 56 –- 61, Aug. 2006.

\bibitem{NetworkMIMO3} G.~J.~Foschini, M.~K.~Karakayali, and R.~A.~Valenzuela, "Coordinating
multiple antenna cellular networks to achieve enormous spectral efficiency," {IEE Proc. Commun.,}
 vol. 153, no. 4, pp. 548 -– 555, Aug. 2006.







\bibitem{hassibi} B.~Hassibi and B.~M.~Hochwald, ``How much training is needed in
multiple-antenna wireless links?'' {\em IEEE Trans. Inf. Theory}, vol. 49, pp.
951–963, Apr. 2003.








\end{thebibliography}
\end{document}